\documentclass[10pt,journal]{IEEEtran}

\makeatletter
\adddialect\l@ENGLISH\l@english
\makeatother

\usepackage{cite}
\usepackage{graphicx}
\usepackage[english]{babel}
\usepackage{longtable}
\usepackage{multicol}
\usepackage{multirow}
\usepackage{epstopdf}
\usepackage[english]{babel}
\usepackage{color}

\begin{document}

\title{Impact of IEEE 802.11n/ac PHY/MAC High Throughput Enhancements over Transport/Application Layer Protocols -- A Survey}
\author{Raja Karmakar, Samiran Chattopadhyay, Sandip Chakraborty
\thanks{Copyright (c) 2016 IEEE. Personal use of this material is permitted. However, permission to use this material for any other purposes must be obtained from the IEEE by sending a request to pubs-permissions@ieee.org.}
\thanks{R. Karmakar is with the Department of Information Technology, Techno India College of Technology Kolkata, INDIA 700156 (E-mail: rkarmakar.tict@gmail.com)}%
\thanks{S. Chattopadhyay is with the Department of Information Technology, Jadavpur University, Kolkata, INDIA 700098 (E-mail: samiranc@it.jusl.ac.in)}%
\thanks{S. Chakraborty is with the Department of Computer Science and Engineering, Indian Institute of Technology Kharagpur, INDIA 781302 (E-mail: sandipc@cse.iitkgp.ernet.in)}%
} 

\markboth{IEEE Communication Surveys and Tutorials}{Karmakar \MakeLowercase{\textit{et al.}}}

\IEEEcompsoctitleabstractindextext{
\begin{abstract}
Since the inception of \emph{Wireless Local Area Networks} (WLANs) in the year 1997, it has tremendously grown in the last few years. IEEE 802.11 is popularly known as WLAN. To provide the last mile wireless broadband connectivity to users, IEEE 802.11 is enriched with IEEE 802.11a, IEEE 802.11b and IEEE 802.11g. More recently, IEEE 802.11n, IEEE 802.11ac and IEEE 802.11ad are introduced with enhancements to the physical (PHY) layer and medium access control (MAC) sublayer to provide much higher data rates and thus these amendments are called \emph{High Throughput WLANs} (HT-WLANs). In IEEE 802.11n, the data rate has increased up to $600$ Mbps, whereas IEEE 802.11ac/ad is expected to support a maximum throughput of $1$ to $7$ Gbps over wireless media. For both standards, PHY is enhanced with multiple-input multiple-output (MIMO) antenna technologies, channel bonding, short guard intervals (SGI), enhanced modulation and coding schemes (MCS). At the same time, MAC layer overhead is reduced by introducing frame aggregation and block acknowledgement technologies. However, existing studies reveal that although PHY and MAC enhancements promise to improve physical data rate significantly, they yield negative impact over upper layer protocols -- mainly for reliable end-to-end transport/application layer protocols. As a consequence, a large number of schools have focused researches on HT-WLANs to improve the coordination among PHY/MAC and upper layer protocols and thus, boost up the performance benefit. In this survey, we discuss the impact of enhancements of PHY/MAC layer in HT-WLANs over transport/application layer protocols. Several measures have been reported in the literature that boost up data rate of WLANs and use aforesaid enhancements effectively for performance benefit of end-to-end protocols. We also point out limitations of existing researches and list down different open challenges that can be explored for the development of next generation HT-WLAN technologies. 
\end{abstract}

\begin{IEEEkeywords}
IEEE 802.11n, IEEE 802.11ac, MIMO, MU-MIMO, Channel bonding, Short-guard interval (SGI), Frame aggregation, Block Acknowledgement, TCP/UDP Throughput.
\end{IEEEkeywords}}

\maketitle

\IEEEdisplaynotcompsoctitleabstractindextext

\IEEEpeerreviewmaketitle

\section{Introduction}

\IEEEPARstart{S}INCE its inception, IEEE 802.11~\cite{standardDot11} or \emph{Wireless Fidelity} (Wi-Fi) based local area networks have gained tremendous popularity and acceptability due to their capabilities of providing the last mile wireless broadband connection with an easy-to-use auto-configuration and self-healing capabilities. Therefore, the development of IEEE 802.11 has witnessed rapid prototyping with a large number of augmentations to the basic protocol. In IEEE 802.11a, IEEE 802.11b and IEEE 802.11g, new PHY features were introduced, whereas new MAC standards were developed in IEEE 802.11e and IEEE 802.11s. Recently, with the extensive demand for capacity and physical data rates, the Wi-Fi technology has been augmented with high to very high throughput extensions like IEEE 802.11n, IEEE 802.11ac and more recently, IEEE 802.11ad.   

{\bf Background:} IEEE 802.11~\cite{standardDot11}, wireless local area network (WLAN), was introduced in $1997$ in $2.4$ GHz frequency band having data rates of $1$ Mbps and $2$ Mbps. IEEE 802.11a and IEEE 802.11b were introduced in $1999$. IEEE 802.11a supports data rate up to $54$ Mbps with frequency band $5$ GHz, whereas IEEE 802.11b supports a maximum of $11$ Mbps. However, the frequency band is lower, which is $2.4$ GHz. In 2003, IEEE 802.11g was introduced with the maximum data rate of $54$ Mbps in $2.4$ GHz frequency band. To define a set of quality of service (QoS) features for WLANs through the modification of MAC, IEEE 802.11e amendment was released in the year of $2005$. However, multimedia applications require more bandwidth for its audio and video transfers and the increasing demand of multimedia applications over WLANs could not be fulfilled by IEEE 802.11a/b/g standards due to their low throughput and low capacity. Hence, there was a requirement to significantly boost up the physical data rate and the overall throughput of IEEE 802.11 based WLANs. Thus, new amendments of WLANs were introduced. IEEE 802.11n~\cite{standardDot11n} was released in $2009$. IEEE 802.11ac~\cite{standardDot11ac} was approved in the year of $2014$. These specifications provided a solution for supporting higher data rate up to $600$ Mbps in IEEE 802.11n and $7$ Gbps in IEEE 802.11ac apart from increasing the range of WLAN coverage. 

To boost up the physical layer throughput, coverage and reliability, IEEE 802.11n~\cite{standardDot11n} and IEEE 802,11ac~\cite{standardDot11ac} include many enhancements. Several modulation techniques and advanced signal processing techniques have been added to the PHY layer exploiting channel bonding and multiple antennas to create wider channels. Some enhancements have been adopted in the MAC sublayer to reduce MAC overhead. The PHY enhancements include MIMO (IEEE 802.11n), multi-user MIMO (IEEE 802.11ac/ad), channel bonding and short guard interval (SGI). Channel bonding increases channel bandwidth by combining multiple $20$ MHz channels. In IEEE 802.11, guard intervals are used to ensure that transmissions belonging to different users do not interfere with each other. In SGI introduced in IEEE 802.11n, $400$ ns guard interval is used in contrast to standard $800$ ns to reduce the interference overhead when only a small number of stations contend for the wireless channel. The HT-WLAN standards also define advanced \emph{modulation and coding schemes} (MCS) -- an assignment of simple integer to every permutation of coding rate, modulation, channel width, SGI and number of MIMO spatial streams. The MAC layer is enhanced by adding \emph{frame aggregation}, \emph{block acknowledgement} (BACK) and \emph{reverse direction} (RD) mechanisms. 

{\bf Existing Surveys and Objective of this Survey:} Although there exists a few surveys on high throughput wireless access networks, like~\cite{Charfi:2013, xiao2005ieee, paul2009evolution}, they primarily discussed about the physical and MAC layer enhancements that help to boost up the physical data rates. However, a large number of recent studies actually deal with the performance analysis of IEEE 802.11n and IEEE 802.11ac networks, and they show that higher PHY and MAC data rates do not necessarily translate into corresponding increments in upper layer, like transport/application layer throughput. For instance, TCP is one of the mostly used reliable transport layer protocols that is used extensively for application development. The recent studies like~\cite{Deek:2011,Deek:2014,Deek:2013} show that TCP indeed exhibits a negative impact on IEEE 802.11n channel bonding; in many cases, TCP goodput at $20$ MHz is found to be more than TCP goodput at $40$ MHz, although $40$ MHz provides better physical data rate. 

This survey focuses on the impact of different enhancements of PHY and MAC layer introduced in IEEE 802.11n and IEEE 802.11ac on upper layer protocols such as TCP and UDP, and on QoS mechanisms. We enumerate the challenges that influence the upper layer performance like transport/application layer throughput using TCP/UDP. We also focus on future works as well as open issues in IEEE 802.11n and IEEE 802.11ac WLANs. In the rest of the paper, we use the terms `upper layer' and `transport/application layer' interchangeably.

{\bf Uniqueness of this Survey:} A survey on PHY/MAC Enhancements with QoS mechanisms for HT-WLANs is presented in~\cite{Charfi:2013}. It elaborately discusses different enhanced PHY and MAC features introduced in several high throughput amendments of IEEE 802.11. The performance of these features is discussed in the presence of QoS and different traffic categories. However, it does not focus on end-to-end performance and thus the impact of several PHY/MAC enhancements over transport/application layer protocols has not been analyzed in that survey. To the best of our knowledge, our survey is the first in this direction to consider the issue of upper layer throughput as a consequence of PHY/MAC enhancements of IEEE 802.11n and IEEE 802.11ac.

{\bf Organization of this Survey:} The rest of the survey is organized as follows. In \S~\ref{sec:over}, we present a brief overview of IEEE 802.11n and IEEE 802.11ac with PHY/MAC enhancements. \S~\ref{sec:group} mentions how the existing \textbf{\textcolor{black}{literature}} has been classified into different tables to present comparative studies, which is followed by an analysis of impact of PHY/MAC enhancements over upper layer protocols. To analyze this impact, we consider one PHY/MAC enhanced feature of HT-WLANs in \S~\ref{sec:one}. Similarly, two, three and all enhanced PHY/MAC features related works are presented in \S~\ref{sec:two},\S~\ref{sec:three} and \S~\ref{sec:all} respectively. In \S~\ref{sec:ieee802.11ac}, we discuss only IEEE 802.11ac related works in this issue. Next, we present possible future directions of research in this domain while highlighting different open research challenges. Possible extensions of existing works as well as open challenges in HT-WLANs are discussed in \S~\ref{sec:n} (focused primarily on the development of IEEE 802.11n), \S~\ref{sec:ac} (focused primarily on the development of IEEE 802.11ac) and \S~\ref{sec:nac} (general research issues in HT-WLANs). Finally, \S~\ref{sec:conc} concludes the survey.


\section{Overview of IEEE 802.11n and IEEE 802.11ac Enhancements}
\label{sec:over}

This section summarizes enhancements over IEEE 802.11n and IEEE 802.11ac for supporting high throughput and very high throughput wireless communication. 

\subsection{Overview of IEEE 802.11n Enhancements}

IEEE 802.11n was released with several enhancements to the IEEE 802.11 PHY and MAC layers~\cite{standardDot11n}. These enhancements significantly improved the performance and reliability of wireless local area network. The main features of IEEE 802.11n can be categorized as -- PHY enhancements and MAC enhancements. A detailed description of several enhanced PHY and MAC features in IEEE 802.11n is presented in the following.

\subsubsection{PHY Enhancements}

\begin{figure}[!ht]
\centering
\includegraphics[width=8.8 cm,height=13.3 cm]{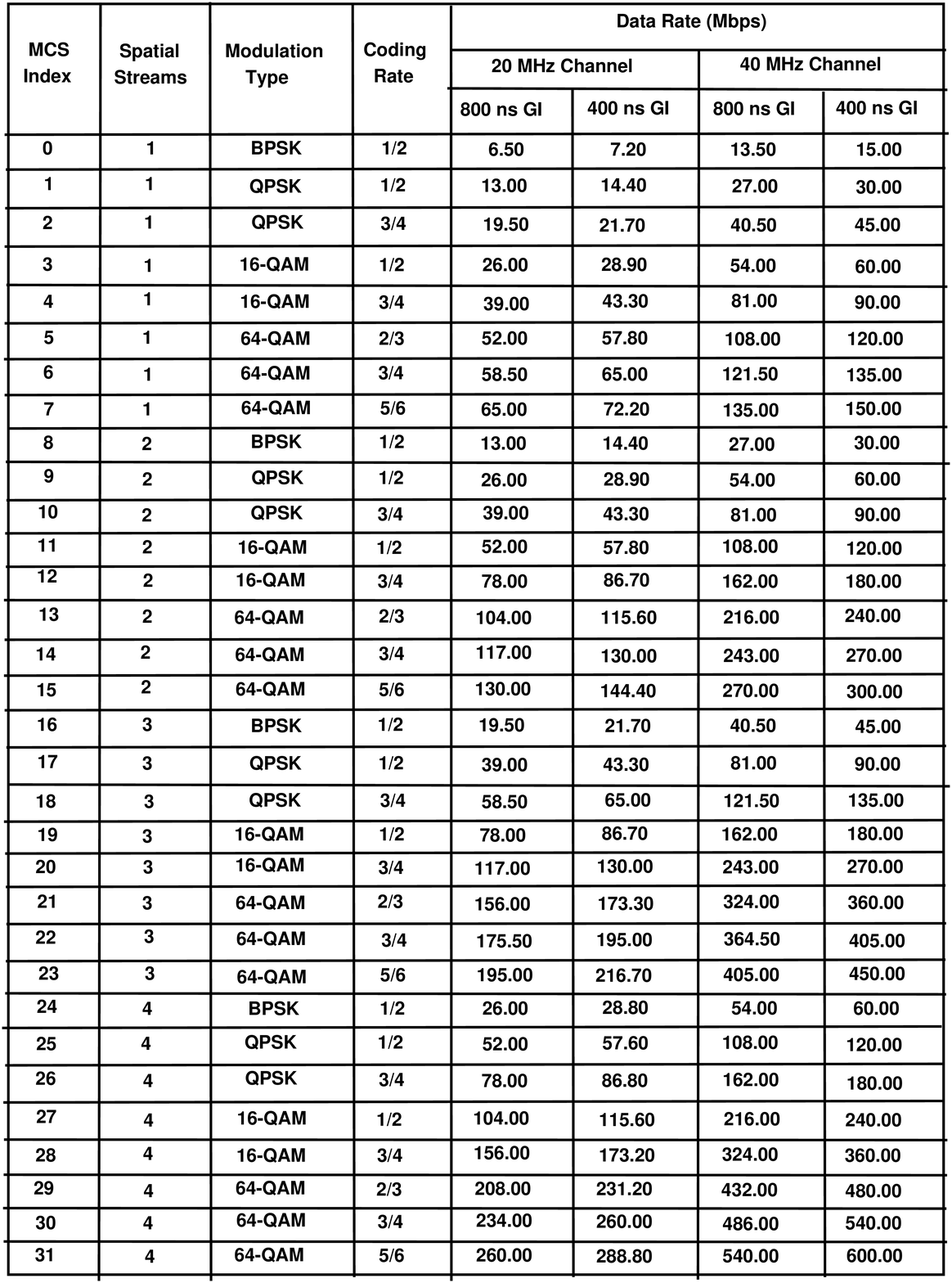}
\caption{MCS Values of IEEE 802.11n}
\label{fig:aggr2}
\end{figure}

There are some enhancements in PHY layer to increase transmission range and data rate. These enhancements are discussed as follows.\\\\
\textit{i) Multiple Input Multiple Output (MIMO):} OFDM-MIMO-based physical layer, which can provide higher data rates and higher range, is introduced in IEEE 802.11n. It is an antenna technology where both the transmitter and the receiver use multiple antennas. The antenna circuits are combined to optimize data rate and minimize errors. Multiple antennas help to transmit and receive data simultaneously. IEEE 802.11n defines $M \times N$ antenna combinations where $M$ is the number of transmit antennas and $N$ denotes the number of antennas in the receiver side. $M$ and $N$ vary from 1 to 4. More than one antenna can be used simultaneously and this mechanism can increase the maximum data rate. There are three types of MIMO mechanisms as given in the following.
\begin{itemize}
\item\textit{Spatial Multiplexing (SM):} In this mechanism, the outgoing signal stream is divided into multiple pieces which are simultaneously transmitted through different antennas. These pieces are called spatial streams which arrive at receiver side with different strengths and delays. It doubles the capacity effectively, and maximizes transmission rate. All access points (APs) must use at least two spatial streams, and stations (STAs) can use as few as one stream.
\item\textit{Space-Time Block Coding (STBC):} Using different coded spatial streams, the outgoing signal streams are redundantly transmitted, each through a different antenna. It helps to improve reliability by reducing the error rate for a given signal-to-noise ratio (SNR). When the number of antennas in the transmitter side is higher than that of the receiver, spatial multiplexing can be combined with this feature.
\item\textit{Transmit Beamforming:} It is a signal transmission technique from an array of multiple antennas to single or multiple users. This feature focuses on the outgoing signal streams at the direction of the intended receiver concentrating on the transmitted Radio Frequency (RF). Hence, it can sustain higher data rates and improve received SNR.
\end{itemize}
There is also an advanced MIMO technology called multi-user MIMO (MU-MIMO) where the available antennas are spread over independent access points each having one or more antennas. In that case, separate signals can be transmitted and received simultaneously by multiple transmitters and receivers respectively in the same band.\\\\
\textit{ii) Channel bonding:} This is a mechanism to deal with combining two adjacent 20 MHz channels to create a 40 MHz channel. Ideally, it can double the PHY data rate, but other devices are left with fewer channels. In 2.4 GHz frequency band, there are three non-overlapping 20 MHz channels. In 5 GHz frequency range, 24 non-overlapping 20 MHz channels and maximum of 12 non-overlapping 40 MHz channels are available. Furthermore, 5 GHz frequency band suffers from less interference since 2.4 GHz band shares frequency with commonly used consumer products like Bluetooth, microwave oven, and cordless phones. With proper implementation, 40 MHz channels are more effective than 20 MHz channels depending on the wireless environment.\\\\
\textit{iii) Modulation and Coding Scheme (MCS):} IEEE 802.11n nodes need to be adjusted with the channel width, the number of spatial streams, guard interval, coding rate and the type of RF modulation to be used. The combination of all these features determines the PHY data rate, ranging from 6.5 Mbps to a maximum of 600 Mbps. MCS is an integer assigned to each permutation of channel width, number of spatial streams, guard interval, modulation type and coding rate. Important factors of MCS are discussed in Figure~\ref{fig:aggr2}. 

\begin{itemize}
\item\textit{Modulation and Coding Rate:} This factor determines how data will be sent over the air. IEEE 802.11 started with binary phase-shift keying (BPSK) and IEEE 802.11a added quadrature amplitude modulation (QAM). Updated modulation methods and coding rates are more efficient and sustainable for higher data rates, but older methods are still supported by HT-WLANs for compatibility with backward standards. To achieve data rate around 600 Mbps, IEEE 802.11n uses 64-QAM with coding rate 5/6. Figure~\ref{fig:aggr2} shows the MCS values used in IEEE 802.11n. From this figure, it can be observed that the maximum data rate is achieved for the MCS 31 using 4 spatial streams and at a coding rate of 5/6. It is also shown that 40 MHz channel with 400 ns guard interval achieves the maximum data rate.
\item\textit{Guard Interval:} It is the time interval between transmitted symbols. Guard interval is needed to avoid inter-symbol interference (ISI). Legacy IEEE 802.11a/b/g devices use guard interval of 800 ns. Longer guard interval (800 ns) results in unwanted idle time in data transmission. IEEE 802.11n devices have an option to use 400 ns guard interval. A Short Guard Interval (SGI) can increase the throughput when there is sufficient symbol separation considering the wireless environment. In large network traffic scenario, SGI leads to more interference and thus, reduces throughput due to its short time interval. 
\end{itemize}

MCS values are good representations of combinations of different features of HT-WLANs. With a given value of MCS, we can easily identify values of parameters relating to data transmissions.

\subsubsection{MAC Enhancements}

There are many MAC enhancements introduced in IEEE 802.11 WLANs to decrease MAC overhead. These enhanced features help to achieve high throughput in WLANs.

\textit{i) Frame Aggregation:} Increasing the transmission rate of PHY alone is not enough to meet the desired MAC throughput (more than 100 Mbps). In legacy IEEE 802.11, the MAC overhead has been partially solved by the transmission opportunity (TXOP) technique introduced by IEEE 802.11e amendment. Frame aggregation is very much helpful in enhancing further channel utilization and efficiency of a wireless station. This mechanism combines multiple data packets arrived from the transport/application layer and forms a larger aggregated data frame. In this way, the overhead of frame header as well as inter frame time is saved. The following types of frame aggregations are available.
\begin{itemize}
\item\textit{Aggregated MAC Service Data Unit (A-MSDU):} It combines multiple Logical Link Control (LLC) packets called MAC Service Data Units (MSDUs) to create a single MSDU called A-MSDU. The aggregated frame contains one MAC header, followed by a maximum of 7935 MSDU bytes. Aggregated MSDUs must belong to the same traffic flow i.e., the same TID (Traffic ID) and must have the same destination and source.
\item\textit{Aggregated MAC Protocol Data Unit (A-MPDU):} After the addition of MAC header to each MSDU, multiple MAC Protocol Data Units (MPDUs) are aggregated to create an A-MPDU. It is created before sending the MSDU (or A-MSDU) to the PHY layer for its transmission. The Transmission Identifier (TID) of each MPDU within the same A-MPDU may differ. An A-MPDU has the maximum size of 65535 bytes.

\begin{figure}[!ht]
\centering
\includegraphics[width=8.7 cm,height=7.5 cm]{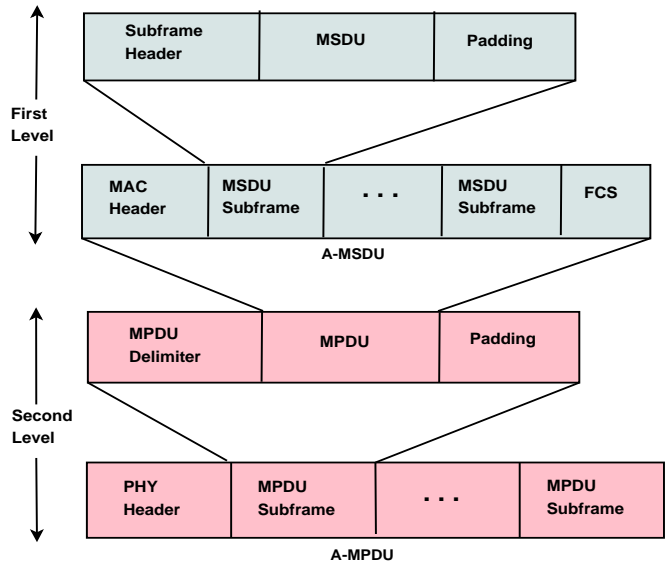}
\caption{Two level frame aggregation}
\label{fig:aggr}
\end{figure}

\item\textit{Multi-Level Frame Aggregation:} A-MSDU and A-MPDU can be combined to produce multi-level frame aggregation. A-MSDU is created in the first level. Then multiple A-MSDUs are aggregated to form a single A-MPDU considering TID, source, destination and the size of A-MSDU. An A-MPDU can only hold complete A-MSDUs or MSDUs. Any fragment of A-MSDUs or MSDUs is not allowed in an A-MPDU. Figure \ref{fig:aggr} demonstrates this concept.
\end{itemize}

\begin{figure}[!ht]
\centering
\includegraphics[width=8.9 cm,height=13 cm]{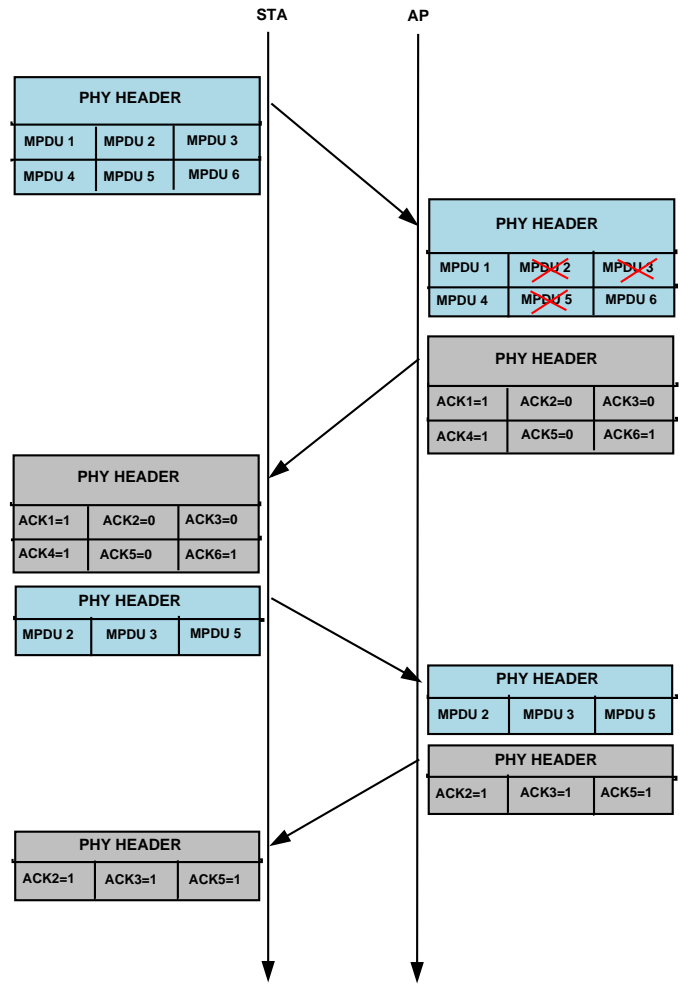}
\caption{Block Acknowledgement}
\label{fig:aggr1}
\end{figure}

\textit{ii) Block Acknowledgement (BACK):} This operation was introduced in IEEE 802.11e that incorporates the TXOP mechanism. This feature is enhanced in IEEE 802.11n to be applied with frame aggregation mechanism as shown in Figure~\ref{fig:aggr1}. An aggregated frame may reduce the overhead significantly in a transmission but frame error rate (FER) is increased with higher size of aggregated frames. Hence, it further reduces the network throughput, and may need multiple retransmissions of the same frame. The BACK mechanism is modified in IEEE 802.11n to support multiple MPDUs in an A-MPDU to overcome this drawback. When the receiver finds errors in some MPDUs of an A-MPDU, it sends a BACK. This BACK contains the acknowledgements of correct MPDUs only. Then, the sender retransmits only the MPDUs which are not acknowledged. This mechanism can be applied to A-MPDU only, but not to an A-MSDU. For error recovery, the whole A-MSDU is needed to be retransmitted when it is found to be incorrect. As the maximum number of MPDUs is 64 in an A-MPDU, one BACK bitmap is able to acknowledge at most 64 MPDUs.

\begin{figure}[!ht]
\centering
\includegraphics[width=8.8 cm,height=7.3 cm]{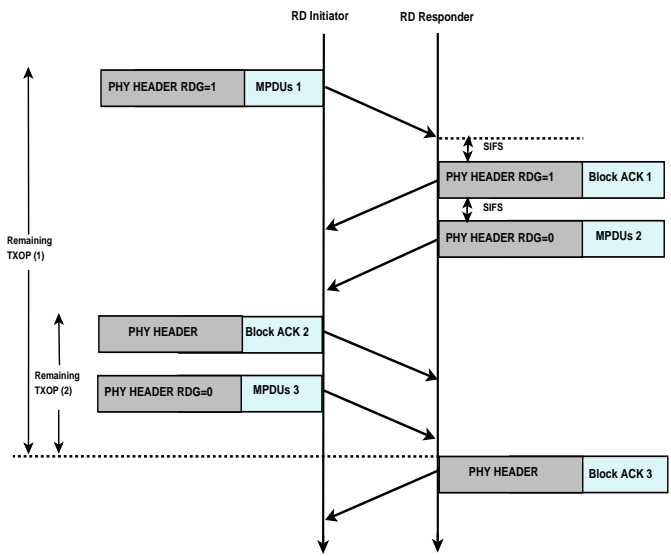}
\caption{Reverse Direction}
\label{fig:aggr3}
\end{figure}

\textit{iii) Reverse Direction (RD):} This feature is responsible for enhancing efficiency of TXOP. In conventional TXOP technique, transmission is unidirectional from the station which holds TXOP. Thus, it is not possible to apply TXOP mechanism in bi-directional network traffic services like Voice over IP (VoIP), video conference, online gaming etc. Hence, performance is degraded by random back off in bi-directional network traffic which leads to lower network throughput. Reverse direction mechanism is shown in Figure \ref{fig:aggr3}. There are two types of wireless stations -- RD initiator and RD responder. After holding TXOP, RD initiator sends a reverse direction grant (RDG) with a data frame to RD responder. After receiving it, RD responder responds with an RDG acknowledgement (ACK) when it has some data to send. If ACK is received, RD initiator waits for the data transmission that will be started from RD responder after a Short Interframe Space (SIFS). SIFS is the time interval between receiving a frame and responding with an ACK frame. If there is more data to be sent, RD initiator can accept or reject the request. Hence, data packets do not wait in the queue until the RD initiator can hold a TXOP. Thus, RD achieves reduction in delay in the reverse link traffic. This feature enhances the channel utilization by allocating the unused time of TXOP to its receivers and thus, the performance of RD is improved.

\subsubsection{IEEE 802.11n Interoperability with Legacy Mode of Wireless Network}

There are millions of legacy IEEE 802.11a/b/g devices which have been deployed and these devices share frequency band used by IEEE 802.11n. This coexistence is a critical issue as old legacy devices cannot be replaced overnight. Hence, IEEE 802.11n should be able to operate effectively with legacy standards with limited impact on legacy WLANs. Also, communication with legacy stations should remain operational. This interoperability can be achieved by the use of High Throughput (HT) protection along with coexistence mechanisms. Therefore, IEEE 802.11n uses three types of modes to work with legacy wireless modes.
\begin{itemize}
\item\textit{High Throughput Mode (Greenfield):} The High Throughput (HT) mode is also called Greenfield mode. It assumes that there are no legacy stations which are using the same frequency band as high throughput stations are using. Hence, no existing legacy station can communicate with the IEEE 802.11n Access Point (AP). This mode is optional. An IEEE 802.11n node with Greenfield mode can communicate only with another node with Greenfield mode.
\item\textit{Non-HT Mode (Legacy):} In this mode, frames can be sent in the IEEE 802.11a/b/g format such that these frames can be understood by legacy stations. For this purpose, the AP must use channels of 20 MHz. To ensure backward compatibility, all products should support this mode. An IEEE 802.11n AP using this mode cannot produce better performance than that of legacy standards.
\item\textit{HT Mixed Mode:} This is the most common operation mode in IEEE 802.11n WLANs. Here HT enhancements and HT protection mechanisms can be used simultaneously to permit communication with legacy stations. HT mixed mode is able to provide backward compatibility but significant throughput penalties are paid as compared to HT mode.
\end{itemize}

\subsubsection{Quality of Service (QoS) for IEEE 802.11n}

Voice and multimedia transmissions on wireless environment require enhanced QoS in IEEE 802.11 MAC standard. IEEE 802.11e is an approved amendment that defines a set of QoS enhancements for WLANs applications implementable through modifications at the MAC. This extension adds QoS facilities with a set of enhancements to improve the performance of WLANs. These new facilities help network managers to configure parameters of networks to match application requirements. Understanding newly added access methods and the effect of their parameters on network performance are important to ensure high throughput in WLANs.

There are two access methods defined in IEEE 802.11 standard. The basic one is {\em Distributed Coordination Function} (DCF) using {\em Carrier Sense Multiple Access with Collision Avoidance} (CSMA/CA). The other method is {\em Point Coordination Function} (PCF). DCF uses an optional {\em Request to Send/Clear to Send} (RTS/CTS) mechanism to share the wireless medium between stations. The sender waits to begin the data transmission until it finds that the medium is idle. The receiver sends an acknowledgement to the sender to inform about the successful transmission of data. When a station senses that the medium is idle, a random number is generated within a given bound defined by {\em Contention Window} (CW). Then, the station begins to decrease the random number. When this number reaches zero, the station starts to transmit data if the wireless medium is free. Otherwise, the station initiates another random number and repeats this procedure. On the other hand, PCF defines a polled protocol but this access mechanism has not yet been widely implemented in WLANs. Both DCF and PCF do not support any type of prioritization in data.

\textit{i) Access methods:} {\em Hybrid Coordination Function} (HCF) is an access method defined by IEEE 802.11e. DCF and PCF are replaced by HCF with new access types to ensure improvement of bandwidth. It also reduces latency of high-priority network traffic and thus, provides high throughput performance. HCF is of two types: \textit{Enhanced Distribution Coordinate Access (EDCA)} and \textit{Hybrid Controlled Channel Access (HCCA)}. EDCA is the extension of DCF and HCCA extends PCF. There are four network access categories (ACs) specified by EDCA. These are Background (BK), Best Effort (BE), Video (VO) and Voice (VO), where each category corresponds to a type of data service. There are four parameters configuring for each of these access categories: CWmax (Maximum Contention Window) CWmin (Minimum Contention Window), TXOP limit and AIFS (Arbitration Inter-Frame Space). 

The configuration of parameters of different classes of data services helps network managers in tuning networks for traffic loads and mix of applications. A station waits for sending data until it finds that the medium is not busy. In EDCA, the station waits for an additional time period which is called AIFS. This additional time period is determined by the type of transmission. The AIFS value of access category of data determines this additional waiting period. It should be a small value for voice data. For other data, a random number is generated between CWmin and CWmax by stations using AIFS. A low CWmax and low CWmin should be used in data with high-priority. The TXOP limit of access category specifies the maximum length of a transmission. If the data is too large, it is split into multiple transmissions. Voice data should have small TXOP since its packets are short in size. HCCA is also a polled protocol like PCF. In HCCA, a station waits for even a shorter time almost comparable to AIFS of EDCA users. HCCA configures QoS settings for each application individually. There is a Hybrid Coordinator (HC) in HCCA mechanism and it is located in AP. HC polls individual stations and grants access to wireless medium based on a specific configured QoS settings. As the network traffic increases, high priority data service does not suffer since there is no contention in the network.

In IEEE 802.11n, QoS better than IEEE 802.11e can be incorporated. EDCA and HCCA provide better performance than legacy QoS. Using these QoS mechanisms, IEEE 802.11n can achieve high throughput.

\textit{ii) Automatic power save delivery:} Automatic Power Save Delivery (APSD) is a new type of delivery having power save mode with notification mechanisms. It was introduced in IEEE 802.11e. It is also incorporated in IEEE 802.11n. As it leads to low power consumption, APSD is a more efficient mechanism of power management than the legacy IEEE 802.11. There are two ways in APSD to start delivery. One of them is called Scheduled APSD (S-APSD) and another one is Unscheduled APSD (U-APSD). In APSD, multiple frames can be transferred together by an AP to a power-saving wireless station during a service period. In S-APSD, service periods are started according to a predefined schedule which is known to the device equipped with power-saving mode. Hence, it allows AP to transfer its buffered data to wireless stations without any prior signaling. In U-APSD, when a frame is transmitted to an AP, a service period is initiated which allows the AP to send buffered traffic in another direction. Both EDCA and HCCA support S-APSD. U-APSD is available only in EDCA. 

\begin{table*}[!ht]
\caption{Comparative study of IEEE 802.11 High Throughput WLAN (IEEE 802.11n and IEEE 802.11ac):}
\centering
\begin{tabular}{|p{2cm}|p{5cm}|p{5cm}|}
\hline 
 & & \\
\textbf{Item} & \textbf{IEEE 802.11n} & \textbf{IEEE 802.11ac}\\ [1ex]
\hline
Release & October, 2009 & January, 2014\\
\hline
Channel & 20/ 40 MHz & 20/ 40/ 80/ 160 MHz\\
\hline
Data Rate & 600 Mbps & 1 Gbps\\
\hline
MIMO stream & $4\times 4$ & $8\times 8$\\
\hline
MAC Mechanisms & Frame Aggregation (A-MSDU and A-MPDU), Block Acknowledgement, Reverse Direction (RD) &  Enhanced Frame Aggregation (large sizes)\\
\hline
PHY layer (Modulation) & BPSK/ QPSK/ 16 QAM/ 64 QAM & BPSK/ QPSK/ 16 QAM/ 64 QAM/ 256 QAM\\
\hline
Frequency Band & 2.4/ 5 GHz & 5 GHz\\
\hline
Maximum Coding Rate & 5/6 & 5/6\\
\hline
Purpose & High Throughput (600 Mbps) & Very High Throughput (approximately 7 Gbps)\\
\hline
\end{tabular} 
\label{table:1}
\end{table*}

\begin{table*}[!ht]
\caption{IEEE 802.11ac Maximum Achievable PHY Data Rates:}
\centering
\begin{tabular}{|*5{p{2.1cm}|}}
\hline
\textbf{No. of MIMO spatial streams} & \multicolumn{4}{|c|}{\textbf{Data Rate (Mbps)}}\\
\cline{2-5}
 & \textbf{20 MHz channel} & \textbf{40 MHz channel} & \textbf{80 MHz channel} & \textbf{160 MHz channel}\\[0.5em]
\cline{1-5}
1 & 86.7 & 200 & 433.3 & 866.7\\
\hline
2 & 173.3 & 400 & 866.7 & 1733\\
\hline
3 & 288.9 & 600 & 1300 & 2340\\
\hline
4 & 346.7 & 800 & 1733 & 3466\\
\hline
5 & 433.3 & 1000 & 2166 & 4333\\
\hline
6 & 577.8 & 1200 & 2340 & 5200\\
\hline
7 & 606.7 & 1400 & 3033 & 6066.7\\
\hline
8 & 693.3 & 1600 & 3466 & 6933\\
\hline
\end{tabular}
\label{table:2}
\end{table*}

\subsection{Overview of IEEE 802.11ac Enhancements}
\label{sec:overac}

IEEE 802.11n was introduced to improve the performance of WLANs compared to legacy standards. It increases the maximum data rate up to 600 Mbps theoretically which is significantly higher than 54 Mbps theoretically achievable by previous standards. For further improvement of network throughput, a new standard, IEEE 802.11ac, was introduced in 2014. Table~\ref{table:1} presents a comparative study of IEEE 802.11n and IEEE 802.11ac. This specification is built on IEEE 802.11n standard that expands channel bandwidth and the number of MIMO spatial streams. 

The PHY layer enhancements of IEEE 802.11ac are discussed in the following.

{\textit{i) Mandatory 5 GHz Band:} The earlier wireless standards (IEEE 802.11b/g/n) operate in 2.4 GHz band and IEEE 802.11a operates in 5 GHz frequency band. Optionally, IEEE 802.11n can also support the frequency band of 5 GHz. But, IEEE 802.11ac can only operate in 5 GHz band. The 2.4 GHz frequency band is susceptible to greater interference due to many household devices which are legacy WLANs. This interference is relatively low in 5 GHz band which has 25 non-overlapping channels. The number of channels is greater than that of 2.4 GHz band which has only 3 non-overlapping channels. Hence, IEEE 802.11ac is expected to be affected by less interference. 

\begin{figure}[!ht]
\centering
\includegraphics[width=8 cm,height=6 cm]{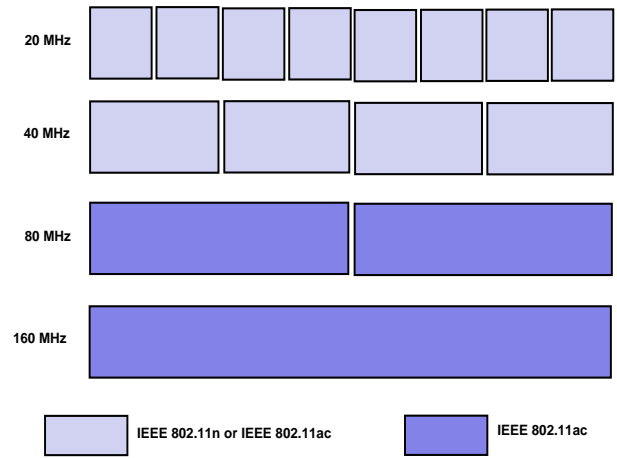}
\caption{Wider Channel Bandwidths in IEEE 802.11ac}
\label{fig:aggr4}
\end{figure}

\textit{ii) Wider Bandwidth:} In addition to 20/40 MHz channel bands, IEEE 802.11ac adds 80 MHz channel band with an optional 160 MHz band as shown in Figure \ref{fig:aggr4}. Channels of 80 MHz can be formed by the combination of two adjacent and non-overlapping 40 MHz bands. Further, 160 MHz channel can be created using two adjacent and non-overlapping channels of 80 MHz frequency band. Thus, IEEE 802.11ac provides more options of channel bonding to achieve better service than legacy standards. Wide range of channel bonding can increase the transmission range.

Higher data rates can be achieved with wider channel bandwidth. Table~\ref{table:2} shows the maximum theoretical data transmission rates in IEEE 802.11ac. This standard considers the highest possible coding scheme and modulation for a given channel bandwidth with a number of MIMO spatial streams. From Table~\ref{table:2}, the maximum transmission rate is seen to be 866.7 Mbps using eight MIMO spatial streams. It is achieved by coding rate of 5/6, Quadrature Amplitude Modulation (QAM) and guard interval of 400 ns. 

As a consequence, IEEE 802.11ac can achieve theoretically even higher throughput than IEEE 802.11n.

\textit{iii) Higher Order MIMO:} Using 1 MIMO spatial stream, IEEE 802.11ac can achieve throughput of at least 1 Gbps in multi-station environments and at least 500 Mbps in single link. In IEEE 802.11ac, the number of MIMO streams is increased up to 8. Thus, a maximum of 8 antennas can be used simultaneously. Hence, more data can be transmitted at a given time interval and thus, network throughput is increased. IEEE 802.11ac can provide a very high throughput (approximately 7 Gbps using 8 MIMO spatial streams) in WLANs using 5 GHz band. Power consumption is also reduced using higher order MIMO.

\begin{figure}[!ht]
\centering
\includegraphics[width=8.2 cm,height=5.5 cm]{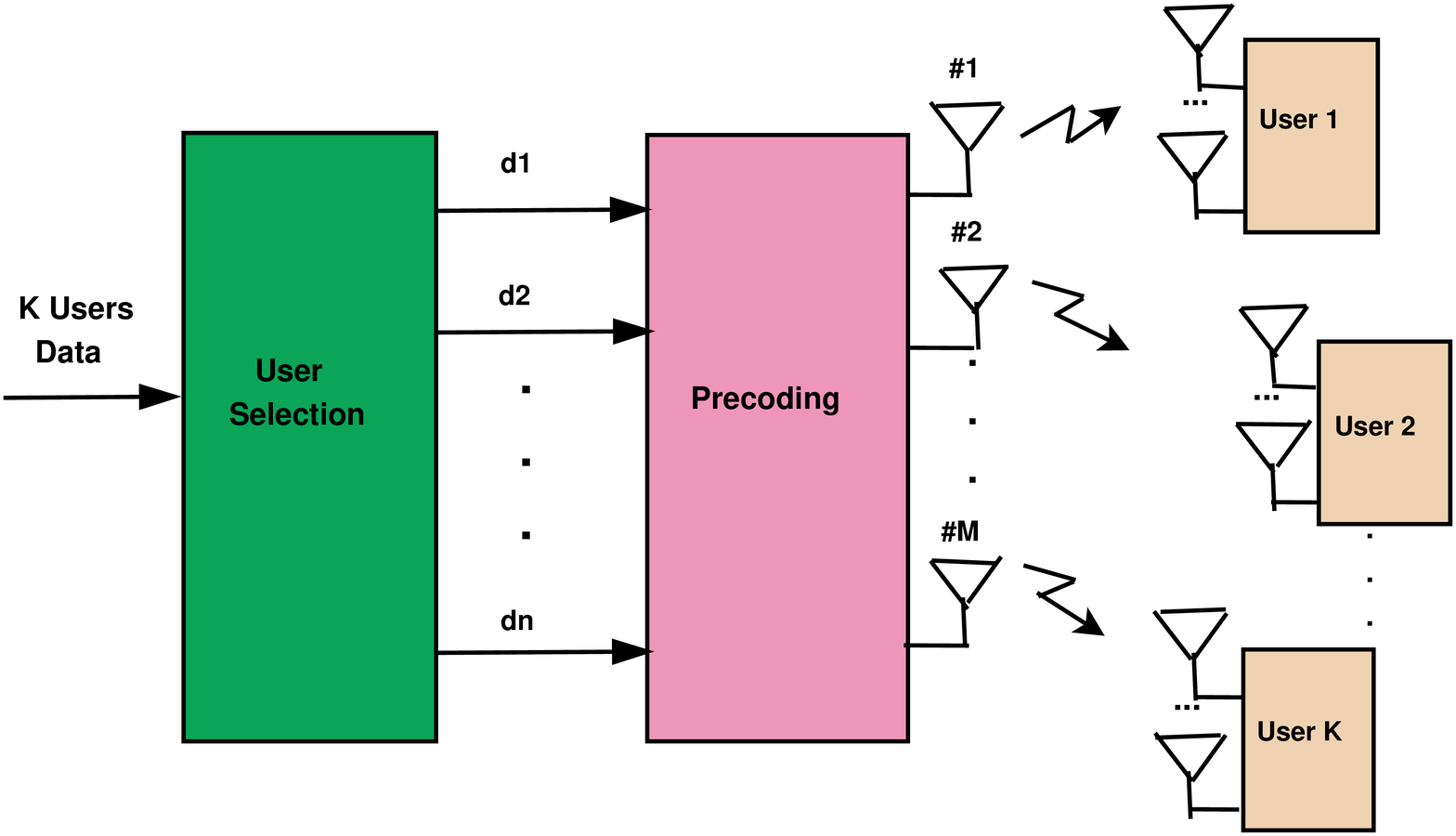}
\caption{MU-MIMO}
\label{fig:aggr5}
\end{figure}

\textit{iv) Multi-User MIMO (MU-MIMO):} The concept of MU-MIMO was introduced first in IEEE 802.11ac. It is considered as an optional feature of IEEE 802.11ac In MU-MIMO, a set of wireless stations having one or more antennas can communicate with each other. A transmitter with multiple antennas can transmit data to multiple receivers simultaneously. Each receiver can have one or multiple antennas. Thus, the transmitter is equiped with MIMO antennas and each of the receivers also has MIMO spatial streams. In this context, multi-user indicates multiple receivers which can be served by a single wireless station at the same time instant. In the transmitter side, each antenna serves as a receiver. As a result, if the transmitter has $n$ number of antennas, it can communicate with a maximum of $n$ number of wireless receivers simultaneously.

Hence, in IEEE 802.11ac, an AP can send multiple packets to multiple stations simultaneously. This can be possible by using up to a maximum of eight MIMO streams. These MIMO streams can be allowed to be divided among a maximum of four stations. Figure~\ref{fig:aggr5} illustrates this mechanism. Each station can use a maximum of four MIMO spatial streams in MU-MIMO transmission mode, supporting up to four clients in the downlink. As MU-MIMO enables an AP to transmit data to multiple stations simultaneously, this feature can greatly increase the throughput of network. Without MU-MIMO, every station may be served one at a time by the AP. Thus, the average throughput effectively is reduced by factor of four. Time-division multiplexing of data is required for Single user MIMO (SU-MIMO) to support multiple stations; and more antennas are needed for reception of data. Hence, SU-MIMO increases device cost.  

\begin{figure}[!ht]
\centering
\includegraphics[width=8 cm,height=6.8 cm]{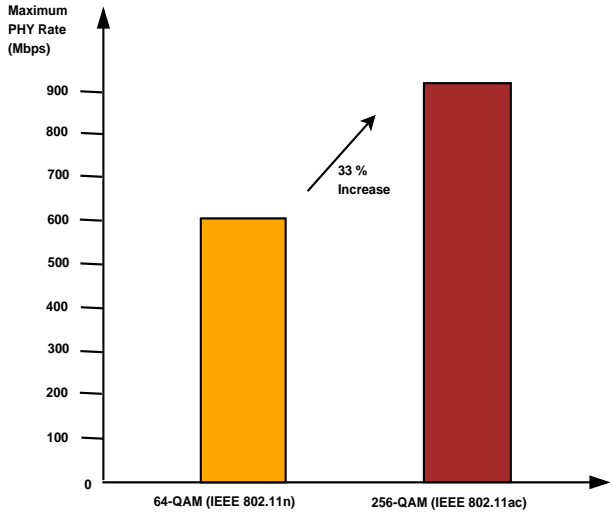}
\caption{Wider Modulation in IEEE 802.11ac}
\label{fig:aggr6}
\end{figure}

\textit{v) Higher Order Modulation:} The highest order of modulation is 64 QAM in case of IEEE 802.11n. In IEEE 802.11ac, the constellation configuration is increased to 256 QAM. This feature increases transmission rate by 33\% compared to IEEE 802.11n as shown in figure~\ref{fig:aggr6}. It is achieved by using eight coded bits/symbol instead of six bits. This higher order modulation helps to achieve very high throughput in WLANs. From Table~\ref{table:2}, it is noted that the maximum data rate in IEEE 802.11ac is 866.7 Mbps that can be achieved with 256-QAM modulation type. The 256 QAM requires higher Signal to Noise Ratio (SNR) than 64 QAM since constellation symbols are placed close to each other in the former case. This type of placing of symbols makes 256 QAM much more susceptible to noise. 

\begin{figure}[!ht]
\centering
\includegraphics[width=8 cm,height=7.2 cm]{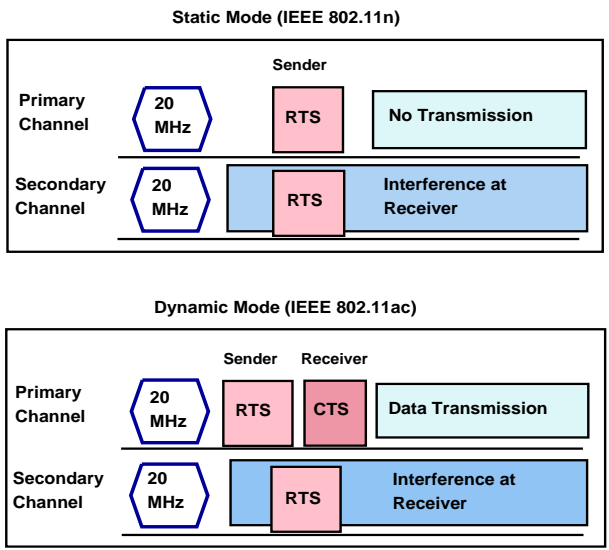}
\caption{Static and Dynamic Bandwidth Management}
\label{fig:aggr7}
\end{figure}

\textit{vi) Dynamic Channel Bandwidth Management:} The management of channel bandwidth is an important issue of any WLANs standard. IEEE 802.11n has available bandwidth of 20 MHz and 40 MHz. IEEE 802.11ac has wider range of bandwidth combinations allowing 20 MHz, 40 MHz, 80 MHz and 160 MHz of channels. Thus, IEEE 802.11ac provides greater flexibility than IEEE 802.11n. The challenge is how to select bandwidth dynamically by optimizing the use of wide bandwidth in an efficient way. Figure~\ref{fig:aggr7} shows static and dynamic bandwidth management. There is a primary channel of 20 MHz which can be accessed by using carrier sensing mechanism, ensuring that there is no interference in networks. Primary channels are also used for backward compatibility and co-existence with previous WLANs standards. Bandwidth can be increased by adding secondary channels. Hence, channel bonding is applied in this regard to create 80 MHz as well as 160 MHz bandwidths from 20 and 40 MHz channels. The set of primary along with secondary channels are managed in wide bandwidth configurations in IEEE 802.11ac. This type of on-demand bandwidth management is known as dynamic channel bandwidth management. RTS and CTS mechanisms of IEEE 802.11 are modified accordingly to improve the operation of this mechanism.

IEEE 802.11n does not properly define CTS/RTS handshake mechanism for managing bandwidth. In static bandwidth management, the receiver sometimes fails to use the actual available bandwidth. At the sender side, there is no interference in primary and secondary channels but the secondary channel has interference at the receiver side. So, the sender sends RTS on both channels but the receiver cannot transmit CTS on either secondary or primary channel in static mode. Hence, no data transmission occurs in this case. 

In dynamic management mode defined in IEEE 802.11ac, channel interference signal is measured as per available channel. Then, the receiver can transmit CTS signal on each channel indicating which channels have no interference. In figure~\ref{fig:aggr7}, the transmission starts on the primary channel in dynamic bandwidth management. Thus, this mechanism improves the overall channel bandwidth utilization as well as network performance.

Mandatory and optional features of IEEE 802.11ac are discussed in Table~\ref{table:3}.

\begin{figure*}
\centering
\includegraphics[scale=0.35]{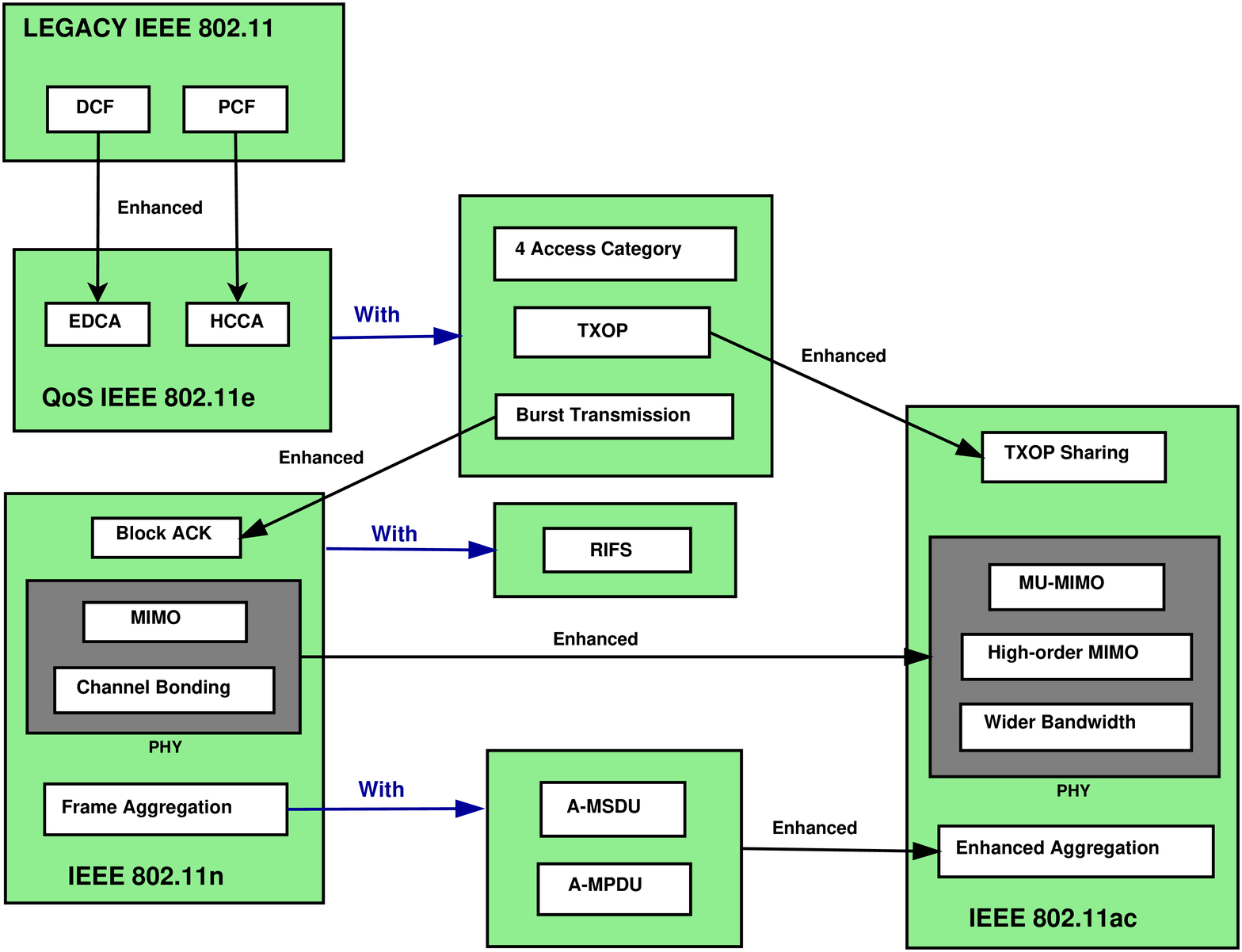}
\caption{All expected MAC and PHY enhancements}
\label{fig:history}
\end{figure*}

\begin{table*}[!ht]
\caption{IEEE 802.11ac mandatory and optional features:}
\centering
\begin{tabular}{|p{2cm}|p{5cm}|p{5cm}|}
\hline 
 & & \\
\textbf{Features} & \textbf{Mandatory} & \textbf{Optional}\\ [1ex]
\hline
Channel bandwidth & 20, 40, 80 MHz & 160 MHz\\
\hline 
Modulation and coding scheme & MCS 0-7 (BPSK 1/2 - 64 QAM 5/6) & MCS 8,9 (256 QAM 3/4,256 QAM 5/6)\\
\hline
Spatial Streams & 1 & 2-8\\
\hline
Beamforming feedback & & Respond to beamforming sounding\\
\hline
Space-time block coding (STBC) & & Transmit and receive \\
\hline
Parity check & & Transmit and receive LDPC\\
\hline
Multi-user MIMO & & Maximum of 4 spatial streams with same MCS per client\\
\hline
\end{tabular} 
\label{table:3}
\end{table*}

\textit{vii) Single Method Closed Loop Transmit Beamforming:} The transmission energy of MIMO spatial streams is focused on target wireless stations by \textit{Beamforming} mechanism. There are several methods regarding beamforming defined in IEEE 802.11n but no one is mandatory for this standard. As a result, chipset vendors implement some non-interoperable beamforming techniques.

IEEE 802.11ac defines a single closed loop mechanism for transmit beamforming. Applying this method, an AP transmits a specific sounding signal to all stations, which estimate channel condition and report the beamforming information (metrices) back to AP. This type of feedback from stations is standardized such that APs and stations from different vendors can interoperate correctly with each other.

In IEEE 802.11ac, transmit beamforming is considered as an optional feature. This mechanism can enable a higher MCS for a given range. Transmit beamforming cannot enhance the maximum range of transmission or increase the maximum data rate. In 5 GHz band, the transmission power is limited by regulatory requirements and thus, the transmit beamforming gain is reduced.\\\\
\textit{viii) Backward Compatibility:} It is required in IEEE 802.11ac to have full compatibility with IEEE 802.11a and IEEE 802.11n. This feature considers a backward compatible preamble which has a section to be understood by IEEE 802.11a/n devices. This property allows legacy wireless devices to work with the IEEE 802.11ac standard as intended.

The MAC layer enhancements of IEEE 802.11ac over IEEE 802.11n are discussed next.


Figure~\ref{fig:history} depicts MAC enhancements in WLANs beginning from legacy standards. It also shows enhanced PHY features in HT-WLANs. IEEE 802.11n introduced two forms of frame aggregation, which are called A-MSDU and A-MPDU. Most of IEEE 802.11n devices implement A-MPDU only. Implementation of both types of frame aggregations has a benefit at data rates over a single type of frame aggregation. As transmission rates of IEEE 802.11ac is much higher, we need to combine both aggregation types to achieve a very high throughput. This combination maintains a good efficiency in overall performance of network. 

In IEEE 802.11ac, the maximum size of A-MSDU has been increased to 11454 bytes. An A-MSDU frame is encapsulated in MPDU. Then, multiple MPDUs are aggregated to create an A-MPDU. The maximum allowable size of a single A-MPDU is up to 1 Mbyte. All packets in IEEE 802.11ac are required to be A-MPDUs. The reason is that PHY signal field does not convey the length of frames in bytes but in OFDM symbols. Furthermore, a single MPDU contains only the packet duration but does not contain any length information. The delimiter in A-MPDU contains the length information of MPDU. Hence, IEEE 802.11ac requires that every packet even if it has just one MPDU, should provide the packet length information in bytes. 

\section{Classification of Existing \textbf{Literature} to Analyze the Impact of IEEE 802.11n and IEEE 802.11ac Extensions over Upper Layer Protocols}
\label{sec:group}

In the earlier sections, we have discussed about the major enhancements introduced in IEEE
802.11n and IEEE 802.11ac. These enhancements are as follows.

\begin{enumerate}
 \item MIMO
 \item Channel Bonding
 \item SGI
 \item Frame Aggregation
 \item BACK (BA)
\end{enumerate}

Additionally, two more PHY features – MU-MIMO and dynamic bandwidth management are introduced in IEEE 802.11ac.

A significant amount of studies on the performance of IEEE 802.11n/ac have been carried out analytically, experimentally and through simulation approaches. These works provide important insights and useful points of reference to assess the impact of PHY/MAC enhancements of HT-WLANs on upper layer performance. This effect on upper layer protocols is manly described as effects on two well established transport layer protocols -- transmission control protocol (TCP) and user datagram protocol (UDP). 

Enhancement of PHY and MAC layers is not sufficient to achieve theoretical high data rates of HT-WLANs in practical scenarios.  Upper layer protocols like TCP, UDP provide end-to-end data transfer. Different layers work cooperatively in networking environment to provide final service to end users. There are many factors which can influence the performance of transport layer in HT-WLANs. Some such important factors are frame aggregation, block acknowledgement, OFDM-MIMO, channel bonding, physical layer capture (PLC) and rate adaptation method. Different link adaptation mechanisms have different impacts on network throughput and thus, they can affect performance of high throughput wireless network.  Efficient power management schemes developed for HT-WLANs may also affect network performance adversely. Therefore, the interaction between the transport layer and PHY/MAC layers is an important area to be explored to achieve high application goodput in practical applications. 

We have summarized works done so far which have discussed about how these enhancements have influenced the physical data rate in the new standards in Table~\ref{table:4} -- Table~\ref{table:16}. Some of these works have also discussed about the impact of these enhancements on upper layer protocols.  We consider performance like TCP/UDP throughput, QoS performance, network capacity, flow performance, network architecture, transmission delay etc. as measures of upper layer protocols. These have been highlighted in the Tables. For each work, we have identified the strengths and weaknesses of the works. A list of such tables is presented in the appendix. 

Many works have considered one or a combination of the aforesaid five/seven features and discussed about their influences on physical data rate and on throughput at upper layers. We have categorised the existing works related to IEEE 802.11n/ac according to the number of enhanced PHY/MAC features considered in the work. Accordingly, works have been summarized in several tables (Table~\ref{table:4} to Table~\ref{table:16}) in the appendix. Table~\ref{table:13} -- Table~\ref{table:16} focus on IEEE 802.11ac related works only in which new PHY features of IEEE 802.11ac are discussed. Table~\ref{table:17} discusses legacy IEEE 802.11 related works which can be extended in IEEE 802.11n/ac for enhancing performance of HT-WLANs.

In all these tables, we denote “channel bonding”, “frame aggregation” and “block acknowledgement” as CB, FA and BA respectively.


\section{Impact of One New Feature of IEEE 802.11n/ac}
\label{sec:one}

Table~\ref{table:4} -- Table~\ref{table:71} focus on existing research works which have considered one new PHY/MAC feature of IEEE 802.11n/ac. These tables discuss the impact of the included feature over upper layer performance. For example, many literature~\cite{Daniel:2012,Adamu:2015,Gong:2012,Anwar:2013,Navid:2011,Vanaja:2013,
Ge:2014,Selvam:2010,Anwar:2012,Marcio:2013} have worked with frame aggregation mechanism only. Only Daniel \textit{et al.}~\cite{Daniel:2012} evaluated upper layer performance using frame aggregation. Dynamic rate adaptation can help HT-WLANs to meet their theoretical throughputs practically. Some existing works have addressed this issue and evaluated transport/application layer throughput considering one enhanced feature of PHY/MAC~\cite{Liang:2014,Chunyi:2015,Xia:2009,Min:2008,Xi:2008,Chi-Yu:2015,Feng:2010,Yun:2010}. 

\subsection{Impact of OFDM-MIMO} 

The MIMO-OFDM is the ideal successor of the present OFDM based wireless networks. Therefore, this feature should be helpful in increasing throughput of WLANs leading to high throughput in applications like wireless video streaming. Table~\ref{table:5} -- Table~\ref{table:7} in the appendix highlight these works. MIMO helps to improve TCP/UDP throughput of HT-WLANs and its performance is analyzed in several literature~\cite{Chen:2006,Paleerat:2014,Mahmood:2013,Lakshmanan:2011,Michael:2014}. OFDM-UWB has a lot of advantages because of its low power consumption and large bandwidth. In~\cite{Chen:2006}, some design parameters of OFDM are highlighted. For example, OFDM-UWB can combat the A/D Converter (ADC) clipping better; it requires less ADC bit precision; it also provides a less Packet Error Rate (PER). Authors in~\cite{Paleerat:2014} focus on trade-offs between MIMO and beamforming technique. Single-User Beamforming (SU-BFM) and Multi-User Beamforming (MU-BFM) provide higher throughput at a longer distance. But, SU-MIMO and MU-MIMO raise data transmission rate at shorter distance. In~\cite{Mahmood:2013}, a reference time synchronization mechanism is applied introducing improved and refined time estimation step. Authors in~\cite{Lakshmanan:2011} discuss that streams and antenna selections are essential to exploit fully the advantages of MIMO technologies. They developed a new metric experimentally for selecting stream. This metric is known as Median Multiplexing Factor (MMF). Rademacher \textit{et al.}~\cite{Michael:2014} showed that the usage of a cross polarized antenna produces a very cost-effective solution for increasing throughput by approximately $100\%$ without changing other QoS parameters like latency. Bandwidth enhancement has the same effect but it suffers from interferences from adjacent links within the frequency spectrum. Table~\ref{table:5} -- Table~\ref{table:7} highlight main features of these works.

The MIMO-OFDM increases network throughput and range of communications significantly due to the use of multiple antennas. MIMO is used in some existing works such as~\cite{Saif:2013,Fang:2009,Ting:2011,Richard:2006} without evaluating TCP/UDP throughput directly. These works can be extended to find impact of MIMO over upper layer throughput. A model is proposed in~\cite{Fang:2009} that emulates the vehicular ratio through the frequency emulation module (FEM) within the signal coverage. This model analyzes the capability of TCP/IP for multiple access point (MAP) mode. Authors in~\cite{Ting:2011} discuss the relationship among different key design parameters such as backhaul link rate, backhaul multihop distance, total coverage size, number of MIMO spatial stream etc. 

Results in~\cite{Richard:2006} show that for a given range, MIMO-OFDM throughput is increased by a maximum of $5$ times than that of non MIMO-OFDM. Experiments show that the use of channel bonding provides a maximum data rate of almost 108 Mbps in 40 MHz frequency band~\cite{Richard:2006}. Hence, IEEE 802.11n can achieve a maximum TCP throughput over 40 Mbps but conventional legacy standards have a maximum TCP throughput of around 25 Mbps. 

An appropriate use of MIMO spatial streams can help in designing an efficient link adaptation scheme. Thus, MIMO can increase end-to-end throughput~\cite{Liang:2014,Xia:2009}. Works~\cite{Liang:2014,Xia:2009} are presented in Table~\ref{table:4} and Table~\ref{table:7}. In~\cite{Liang:2014}, dynamic rate adaptation in MU-MIMO WLANs is decomposed to estimate two values. One of these valuse is the SNR, when wireless station transmits data to AP and the another one is the direction of this transmitted signal received at the AP. The adaptation with the appropriate selection of MIMO mode and MCS value together can achieve more efficient utilization of channel~\cite{Xia:2009}. ESNR is a rate adaptation algorithm proposed in~\cite{Halperin:2010} using MIMO, specially for IEEE 802.11n. This scheme finds the highest rate configuration by applying the current channel state information. The selected configuration is predicted to deliver packets successfully with packet receive rate (PRR) greater than $90\%$. The evaluation of ESNR in~\cite{Halperin:2010} shows that it exploits MIMO feature of IEEE 802.11n and improves the performance of TCP/UDP. Thus, MIMO-OFDM technologies can help to improve the performance of upper layer protocols.

\subsection{Impact of channel bonding} 

The impact of PLC and channel bonding on the performance of wireless mesh network producing high throughput is evaluated in~\cite{Sandip:2015}. Work of~\cite{Sandip:2015} is summarized in Table~\ref{table:4}. It shows that PLC has a tendency to produce negative effect over high throughput mesh WLANs in an asymmetric channel interference (between 20 MHz and 40 MHz channel). When two communications belong to different channel bondings interfering each other, the transmission through 40 MHz channel is lost. And it leads to produce several unfairness in WLANs. In~\cite{Sandip:2015}, it is shown that PLC can improve the upper layer throughput for both links of 20 and 40 MHz in case of 20/40 network scenario. However, in 40/20 MHz condition, PLC is able to improve the throughput significantly only in 20 MHz channel where the overall throughput for 40 MHz channel drops significantly due to the effect of interference of the 20 MHz channel. The 40 MHz channel fails to provide high throughput since it has larger channel width. The network may have symmetric channel interference which means that both channels belong to either 20 or 40 MHz band. In this case, both channels produce long term upper layer throughput fairness following the EDCA air-time fairness mechanism of IEEE 802.11.

\subsection{Impact of frame aggregation} 

The interaction of MAC with TCP is very important since TCP determines process-to-process (end-to-end) transmission rate between the sender and the receiver. Frame aggregation reduces MAC layer overhead because all subframes within an aggregated frame have a common header and trailer. Hence, frame formation time is reduced in MAC. This increases MAC layer throughput due to which the chance of increase of transport/application layer performance is higher. The end-to-end transmission delay decreases as the aggregation size increases since more frames can be delivered within a given period of time in the form of subframes in an aggregated frame. Throughput also becomes saturated after some value of aggregation size because a huge number of subframes will also increase frame error rate.  An appropriate mixture of A-MSDU and A-MPDU need to be considered to reduce MAC overhead and increase data rate~\cite{Daniel:2012} as discussed in Table~\ref{table:6}. This work presents interactions among IEEE 802.11n, the current QoS and power saving mechanisms. It finds that IEEE 802.11n MAC aggregation mechanisms perform better if combined with original power save mode than with IEEE 802.11e U-APSD mechanism. This work also designs an algorithm using information available only at MAC layer to adapt the amount of IEEE 802.11n aggregation according to the level of congestion in network.

The performance of different existing rate adaptation algorithms is evaluated in~\cite{Mohamed:2014} presented in Table~\ref{table:7}, which shows that the reduction of MAC overhead influences link adaptation mechanism. As a result, transport layer throughput is increased. This work discusses two categories of rate adaption algorithms, which are based on collision in the network. The categorization is made according to the metric selected by the sender and/or receiver to calculate the appropriate bit rate for data transmission. 

The appropriate selection of size of frame aggregation helps to select dynamically the number of frames to be transmitted and thus, reduces frame loss ratio. As a result, it becomes an efficient factor for dynamic rate adaptation~\cite{Feng:2010,Chunyi:2015,Selvam:2010} presented in Table~\ref{table:5} and Table~\ref{table:7}. In~\cite{Feng:2010}, the optimal payload size of frame and MCS value are jointly considered according to SNR of channel to maximize network goodput. Results show that the upper layer throughput is enhanced.

Latency is increased due to a long tail distribution of packet delay. Li \textit{et al.}~\cite{Chunyi:2015} proposed a new latency-aware data rate adaptation mechanism that minimizes the tail latency for packet delay-sensitive applications. It considers rate control, software/hardware retransmission dispatching and frame aggregation scheduling. This technique reduces end-to-end transmission delay significantly.


\section{Impact of Two New Features of IEEE 802.11n/ac}
\label{sec:two}

Some existing works have considered any two enhanced features of HT-WLANs. In Table~\ref{table:8} and Table~\ref{table:9}, we mention such works. Few of them have considered the impact of these features on upper layer protocols. Block acknowledgement mechanism is enhanced in~\cite{Nakajima:2007,Nakajima:2005} without considering TCP/UDP performance. Some papers discuss frame aggregation method with BACK such as~\cite{Kolap:2012,Baba:2012,Mustafa:2012,Abu:2011,Santos:2011}. Anwar \textit{et al.}~\cite{Anwar:2014} designed a framework for packet loss differentiation with frame aggregation and BACK. Performance of IEEE 802.11n is analyzed in multi-hop environment using PHY enhancements~\cite{Kwong:2012} and MAC enhancements~\cite{Frohn:2011}. Deek \textit{et al.}~\cite{Deek:2013} proposed a rate adaptation technique considering MIMO and channel bonding. A sender-side link adaptation algorithm based on received signal strength indicator (RSSI) is designed in~\cite{sample:2015}. An efficient channel bonding technique is proposed in~\cite{Deek:2014}. Performance of IEEE 802.11n is examined by some research works~\cite{Wang:2009,Letor:2012}. Channel bonding is an important PHY enhancement and thus, multiple bonding levels can be present simultaneously. Yoriko \textit{et al.}~\cite{Yoriko:2007} discussed a mechanism to control coexistence of $20/40$ MHz bandwidth. At high data rates, link quality is investigated in~\cite{Pel:2010} focusing on MIMO and channel bonding. Authors in~\cite{Xu:2008} discuss highly integrated MIMO, whereas performance of MAC with MIMO is evaluated in~\cite{Alsahlany:2011}. Channel sensing is also necessary for appropriate channel bonding and it is investigated in~\cite{Yao:2012}.

\subsection{Impact of MIMO and channel bonding} 

Wider channel bandwidth cannot practically increase the transmission rate according to the theoretical values. Some recent studies~\cite{Deek:2014,Deek:2013} have proved that the performance of channel bonding with MIMO in IEEE 802.11n is influenced by some factors like interference of neighboring channels and loss of signal strength. Wider channel reduces the range of transmission. Therefore, channel management mechanism in HT-WLANs must recognize the behavior of different types of channel bonding levels. On the basis of this understanding, intelligent channel bonding decisions can be made efficiently~\cite{Deek:2014}. Works~\cite{Deek:2014,Deek:2013} are summarized in Table~\ref{table:8}. In~\cite{Deek:2011}, an evaluation of the impact of varying Packet Receive Rate (PRR) levels on performance is carried out. This work analyzes characteristics of channel bonding in IEEE 802.11n networks and examines factors that influence the behavior of such bonding. The motivation is that a system can predict the behavior of wireless network to maximize its performance. 

Deek \textit{et al.}~\cite{Deek:2014} proposed an intelligent channel bonding scheme based on the transmitter's knowledge of surroundings like signal strength of the channel, interference patterns, channel utilization etc. By these findings, a system can combine multiple 20 MHz channels intelligently to create 40 MHz, 80 MHz and 160 MHz channels. It also calculates channel bonding throughput gain. The experiments also include UDP results for same links to compare against TCP performance. These results show that both TCP and UDP throughputs are increased in 40 MHz channel compared to 20 MHz channel. Hence, data rate may not be doubled in practical situation, where TCP is more sensitive to packet losses. The channel bonding technique used for these experiments is called intelligent channel bonding. Intelligent bonding enhances TCP/UDP throughput significantly~\cite{Deek:2014}. In~\cite{Deek:2013}, a MIMO based rate adaptation scheme is developed and it increases TCP/UDP throughput effectively considering SNR of the channel, PRR, MCS and channel bandwidth. 

There are some parameters which affect performance of channel bonding and thus, affect transport layer throughput~\cite{sample:2015}. These parameters are RSSI values, signal scattering, MCS values etc. These factors should be considered before performing channel bonding to achieve high throughput. In~\cite{sample:2015} discussed in Table~\ref{table:8}, a dynamic link adaptation mechanism is designed. This is a pure RSSI-threshold based scheme. It minimizes sampling overhead and increases the goodput of the system. 

Channel bonding allows to double the PHY rate directly from a 20 MHz channel but MAC and transport/application layer throughputs also depend on other influencing factors (interference, data loss etc.).~\cite{Yoriko:2007} reports that the rate of a channel bonding mechanism used to create a 40 MHz channel depends on positions of 20 MHz channels used for bonding. This work controls the coexistence of 20 MHz and 40 MHz stations by setting Network Allocation Vector (NAV) to these bands in MAC layer. Authors also explained a coexistence proposal for wireless systems where transmission is carried out in both 20 MHz and 40-MHz channels. They also show that the signal strength of neighboring links has great impact on bonded channels because of interference from channel leakage. This interference reduces the performance of 40 MHz channel. Hence, the overall network throughput degrades.

MIMO can boost up throughput significantly under interference, signal fading and multi-path long distance communications. Channel bonding helps to increase physical data rate. Hence, these two features can impact over upper layer performance collectively and thus, increase end-to-end throughput. Impact of MIMO and channel bonding is analyzed for multihop wireless network ~\cite{Kwong:2012}. This work is mentioned in Table~\ref{table:9}. Kwong \textit{et al.}~\cite{Kwong:2012} presented the performance analysis of different parameters like distance between hops, choice of MCS value, multihop chain distance, backhaul link rate, number of MIMO spatial streams, channel bandwidth etc. These features are required to support specific capacity for each AP and connection rate per station. This work finds that at least 3 spatial streams with 16 QAM and coding rate of 3/4 are required to enable 2 Mbps of connection rate per station per AP following the multihop chain. 

Authors in~\cite{Pel:2010} investigated whether high data rates can translate to high link quality in real deployment. In an indoor wireless set up of testbed, it shows that the highest data transmission rate advertised by IEEE 802.11n typically produces losses even in an interference-free environment. Such losses are more acute and can persist at high values of SNR, even at low intensity of interference. It finds that exploitation of spatial diversity and cahnnel bandwith jointly with packet size adaptation can enhance end-to-end performance by reducing packet losses.

\subsection{Impact of frame aggregation and block acknowledgement} 

Frame aggregation and BACK can increase MAC throughput because they reduce MAC overhead. Therefore, TCP/UDP throughput may increase as well. Interaction between TCP and MAC frame aggregation with BACK is evaluated in~\cite{Mustafa:2012}. It focuses on the interaction between frame aggregation and TCP through simulations of different TCP congestion control mechanisms such as Reno, NewReno, Vegas, Selective Acknowledgments (SACK) and Tahoe. It is shown that TCP throughput is influenced by different aggregation sizes. Performance evaluation of high throughput MAC is carried out with frame aggregation and BACK~\cite{Santos:2011}. This work shows that frame aggregation can reduce MAC overhead delay effectively by eliminating header for each subframe. Whereas BACK reduces individual acknowledgement transmission overhead. Hence, these two features can be merged efficiently to reduce MAC processing delay and thus, help to enhance upper layer throughput. This work is mentioned in Table~\ref{table:9} in the appendix.

Frohn \textit{et al.} \cite{Frohn:2011} showed that when several stations send data simultaneously, frame aggregation with BACK reduces each station's MAC overhead. Thus, FA with BACK can increase upper layer performance of overall network. It presents MAC throughput as a function of aggregation level, bit error rate (BER) and path length. This work demonstrates that current TCP implementation does not harness the channel bandwidth mechanism provided by PHY layer of IEEE 802.11n.


\section{Impact of Three Enhanced Features of IEEE 802.11n/ac}
\label{sec:three}

There are very little research that considered three new PHY/MAC features of HT-WLANs to assess their impact on upper layer performance. Table~\ref{table:10} discusses few works that consider three PHY/MAC enhancements. Using MIMO, channel bonding and frame aggregation, basic performance of IEEE 802.11n is analyzed in~\cite{Janis:2012}. Throughput analysis of IEEE 802.11n is presented with MIMO, channel bonding and BACK~\cite{Ghaleb:2012}. Analysis of VoIP for high throughput transmission is discussed in~\cite{Biswas:2007}. Sandip \textit{et al.}~\cite{Sandip:2014} evaluated the TCP performance in burst and correlated losses in high data rate communication. 

\subsection{Impact of channel bonding and frame aggregation with BACK} 

The work in~\cite{Sandip:2014} discusses the impact of channel bonding, frame aggregation and BACK on TCP using Minstrel rate adaptation algorithm in mesh wireless networks. It uses four TCP variants: Loss Tolerant TCP (LT-TCP), Network Coded TCP (TCP/NC), TCP-Horizon and Wireless Control Protocol (WCP). TCP/NC shows the poorest performance among the four TCP variants using 20 MHz channel. LT-TCP improves TCP performance compared to TCP/NC at 40 MHz channel under low MCS level. But, TCP-Horizon and WCP improve transport protocol performance over the other two transport protocol variants. These two protocols improve average goodput by balancing available capacity over all flows. One of the important observations is that WCP performance decreases at higher MCS levels in 40 MHz channel and the same also happens for TCP-Horizon. Loss sensitive TCP variants like TCP/NC and LT-TCP outperform WCP and TCP-Horizon at higher MCS levels in 40 MHz channel. The best transport layer protocol variant in a high throughput mesh network depends on the selection of MCS level. It is very difficult for a transport layer protocol to perform well on a multi-MCS high throughput mesh network. One important point is that end-to-end user-level goodput for all TCP variants is typically at most one half of the maximum network capacity when data rate is more than 150 Mbps. WCP performs poorly at 40 MHz channel over 150 Mbps data rate though it performs better than other protocol variants at 20 MHz channel~\cite{Sandip:2014}.

\noindent{\textbf{Impact of Channel Bonding on WCP performance:}} Channel bonding introduces high error during transmission. In a mesh network, this error propagates to transport/application layer and causes lots of retransmissions in transport layer due to additional noise. Losses in WCP over 40 MHz are due to combined effect of channel errors, network interference and congestion. Channel bonding with dense MCS values shows a negative impact on WCP. Sandip \textit{et al.} \cite{Sandip:2014} showed the impact of frame aggregation and BACK on WCP performance along with channel bonding. In~\cite{Sandip:2014}, it has been shown that A-MSDU performs better than A-MPDU where WCP goodput improves more in A-MSDU, compared to A-MPDU. WCP goodput shows a negative impact for A-MPDUs over 40 MHz channel with dense MCS values. The evaluation of WCP behavior over an IEEE 802.11n+s mesh testbed provides two important observations~\cite{Sandip:2014}. First, the current implementation of IEEE 802.11n channel bonding and frame aggregation mechanisms over mesh network is not suitable for supporting high data rates (at 40 MHz channel with dense MCS values). The second observation is that the problem of WCP over 40 MHz with dense MCS cannot be solved only at the transport layer since PHY and MAC layers greatly affect transport layer. To gain theoretically achievable throughput, transport layer should also be enhanced along with enhancements of PHY and MAC layer.

\subsection{Impact of MIMO, channel bonding and frame aggregation} 

MIMO can boost up throughput significantly under interference, signal fading etc. Channel bonding helps to increase physical data rate. Thus, if frame aggregation that reduces MAC overhead is applied in conjunction with MIMO and channel bonding, TCP/UDP throughput can be further enhanced. As a result, upper layer throughput may be increased in HT-WLANs~\cite{Janis:2012}. Authors of~\cite{Janis:2012} (Table~\ref{table:10}), evaluate the performance of both IEEE 802.11g and IEEE 802.11n considering a wireless communication scenario in an outdoor experiment. This work shows that distance, SNR of the channel and throughput characterize basic standards of upper layer performance. As a result, TCP/UDP throughput can be enhanced.

\subsection{Impact of MIMO, channel bonding and BACK} 

BACK reduces the overhead of transmission of acknowledgement frame by sending multiple acknowledgements in a single acknowledgement frame and thus, increases MAC throughput. Combining BACK with MIMO and channel bonding can further help to enhance MAC throughput. Such combination has a positive impact on TCP/UDP throughput~\cite{Ghaleb:2012}. The performance of IEEE 802.11n is evaluated using a well-known OPNET simulator in~\cite{Ghaleb:2012} combining MIMO, channel bonding and BACK. This evaluation considers influence of different types of access categories on the maximum throughput and the achieved Throughput Efficiency (TE). It also shows that higher MCS values can help to improve TCP/UDP throughput. This work is discussed in Table~\ref{table:10}.

\subsection{Impact of channel bonding, SGI and frame aggregation} 

In~\cite{karmakar2015dynamic}, discussed in Table~\ref{table:10}, a dynamic link adaptation mechanism is proposed for HT-WLANs. In this algorithm, channel bonding, SGI, frame aggregation and different MCS levels are combined to construct the configuration set for data transmission. A state transition model is also used to select the best possible data rate. This selection is carried out on the basis of the past information stored in a statistic table. Results show that this proposed algorithm produces significantly better performance in terms of TCP throughput, packet loss ratio and delay than other competing mechanisms.


\section{Impact of All New Features of IEEE 802.11n/ac}
\label{sec:all}

Very few works have considered all enhanced features of PHY/MAC in HT-WLANs and a few of them have analyzed upper layer performance. In the appendix, Table~\ref{table:11} and Table~\ref{table:12} show the recent literature which have included all PHY/MAC enhancements of IEEE 802.11n/ac. PHY enhancements mainly increase physical data rate in spite of the presence of interference, signal fading etc. Whereas, MAC enhancements reduce MAC processing delay and increase MAC throughput. Hence, the combination of PHY and MAC enhanced features can increase upper layer performance as studied in several existing works~\cite{Funa:2012,Zawar:2015,Sandip:2016}. In~\cite{Funa:2012} (Table~\ref{table:12}), throughput measurement results are analyzed using commercial products which implement the IEEE 802.11n protocol. There is quick drop of end-to-end throughput as the link distance increases. This work also presents a routing tree algorithm minimizing routing delay designed for Wireless Internet-access Mesh Network (WIMNET). The implementation of the algorithm considers the change of link speed by the distance. 

In the presence of external interference, experiments in real indoor environment are carried out and they quantify the gain in the average upper layer throughput provided by IEEE 802.11ac. The gain in throughput is compared with IEEE 802.11n~\cite{Zawar:2015}. This work is presented in Table~\ref{table:12}.

Sandip \textit{et al.} \cite{Sandip:2016} as given in Table~\ref{table:11} proposed an estimation with sampling scheme to filter out features which are non-preferable. Then, the best configuration features are found dynamically by applying a learning mechanism. It is based on Kalman filtering mechanism which finds the preferable parameter sets from all combinations of possible feature sets. The proposed mechanism improves TCP/UDP throughput significantly considering all PHY/MAC enhancements.

Performance of multimedia multicast transmission requiring high channel bandwidth and low transmission delay is analyzed in~\cite{Letor:2012} mentioned in Table~\ref{table:11}. In this case, upper layer throughput is enhanced with PHY and MAC enhancements. 

Charfi \textit{et al.}~\cite{Charfi:2013} presented a survey of QoS and PHY/MAC enhancements of HT-WLANs. This survey discusses elaborately new features of HT-WLANs and their impacts over QoS performance of upper layer protocols. It is mentioned that QoS performance is enhanced by introducing several PHY/MAC enhanced features of HT-WLANs. 

The work~\cite{Nobuo:2013} in Table~\ref{table:11} designs a modified active AP selection scheme and it impacts positively over TCP/UDP performance. In this work, authors propose a modification of the existing active AP selection mechanism to consider the change of link speed. This work also introduces a new design parameter set for WIMNET to choose links having sufficient throughputs to enhance end-to-end throughput effectively.

In Table~\ref{table:12}, routing algorithm for IEEE 802.11n is mentioned in~\cite{Toru:2013,Funa:2012}. In this algorithm, a large speed change is considered in order to cope up the IEEE 802.11n network in WIMNET efficiently. In addition to links between APs, the modified algorithm selects links between hosts and their associated APs. The proposed routing mechanism is able to reduce the routing and AP association delay, which in turn helps to boost up TCP/UDP throughput as mentioned in Table~\ref{table:12}.  

Considering link speed in mesh networks, clustering algorithm is also extended~\cite{Tajima:2013}. A clustering algorithm is proposed in this work for composing a large scale of WIMNET efficiently by partitioning APs into a defined set of clusters. For simplicity, this proposed algorithm assumes that link speed remains constant. Whereas, their preliminary experiments found a great decrease of link speed as the link distance increases. It is due to interference in channel. In the proposed mechanism in~\cite{Tajima:2013}, the use of clusters helps to decrease transmission delay. Thus, end-to-end throughput can be improved and it is discussed in Table~\ref{table:12}.


\section{Works that Considered Impact of Only IEEE 802.11ac Related Works}
\label{sec:ieee802.11ac}

In the appendix, Table~\ref{table:13} -- Table~\ref{table:16} focus only on existing IEEE 802.11ac related research works including new PHY features (MU-MIMO and dynamic channel management). PHY/MAC enhancements improve PHY and MAC performance. It helps to increase TCP/UDP throughput even under weak signal strength and high interference scenario since new features increase sustainability of wireless station in dynamic channel condition. Some of them have evaluated upper layer performance~\cite{Dia:2014,Liao:2013,Faridi:2014,Park:2011,Go:2010,Per:2014,Hart:2011,
Swetank:2015,Prasant:2014,Rung-Shiang:2015,Okhwan:2015,Yazid:2014,Zhiqun:2015,Narayan:2015}. 

\subsection{Impact of MU-MIMO}

MU-MIMO is an important feature of IEEE 802.11ac. Its performance from different perspectives is discussed in~\cite{Oh:2014,Khavasi:2014,Esslaoui:2012}. Authors in~\cite{Oh:2014} suggest two transmission schemes working cooperatively to reduce interference in overlapped basic service set (BSS) of IEEE 802.11ac broadcasting network. These schemes are interference alignment (IA) and MU-MIMO with Time Division Multiple Access (TDMA). MU-MIMO with TDMA mechanism requires lower overhead compared to IA. This is because TDMA-based MU-MIMO considers feedback of channel state information (CSI) from AP to station and AP polling transmission. This approach determines the data transmission order of MU-MIMO with TDMA. Thus, the proposed mechanism improves overall network capacity. 

\noindent{\textbf{Fairness in performance:}} Fairness of IEEE 802.11ac network is examined in some existing works. Khavasi \textit{et al.}~\cite{Khavasi:2014} formulated a simple model to design the max-min fair related link adaptation problem. This model guarantees the minimum gain of utility of each receiver node according to its requirement and thus, upper layer performance is improved. The design of a fair multiuser proportional scheduling mechanism is proposed in~\cite{Esslaoui:2012}, which is based on MU-MIMO OFDM. This scheduling scheme is capable of scheduling simultaneous data transmissions to several users and provides a high value of fairness at application level. This mechanism also enables a wireless system to select different transmission parameters appropriately. As a result, user level throughput increases. Table~\ref{table:15} and Table~\ref{table:16} highlight these works.

\subsection{Impact of MU-MIMO and channel bonding}

Measurements reported in~\cite{Dia:2014} (Table~\ref{table:13}) were carried out in an office building. This work shows that IEEE 802.11ac provides significantly better performance than IEEE 802.11n for small distances. These improvements are found to be sensitive to different channel conditions. It is also observed that channel interference severely impacts upper layer performance in an adverse manner. Sur \textit{et al.}~\cite{Sur:2016} designed MUSE which is a MU-MIMO-based rate selection scheme for IEEE 802.11ac commodity networks. This mechanism uses compressed beamforming feedback from IEEE 802.11ac users and identifies MU-MIMO groups providing the best throughput. MUSE also adjusts channel bandwidth and increases MU-MIMO grouping opportunities. In this work, the evaluation shows that TCP and UDP throughputs are gained significantly over competing schemes mentioned in the work. This is due to the use of MU-MIMO and channel bonding jointly on the basis of beamforming feedback. The impact of MUSE on TCP/UDP throughput is summarized in Table~\ref{table:MUSE}~\cite{Sur:2016}. Authors in~\cite{Sur:2016} compared TCP/UDP throughput gains with respect to Legacy-User Selection (Legacy-US) and PUMA~\cite{NarendraPUMA:2015}. From Table~\ref{table:MUSE}, we can observe that application of MU-MIMO with channel bonding can increase upper layer throughput effectively.

\begin{table}[!ht]
\caption{TCP/UDP throughput gains of MUSE in various settings}
\centering
\begin{tabular}{|p{1.4cm}|p{1cm}|p{1cm}|p{1.5cm}|p{1.5cm}|}
\hline 
Algorithm comparison & Upper layer protocol & Gain Distribution & Aggregate throughput gain (\%) in static condition & Aggregate throughput gain (\%) in dynamic condition\\ [1ex]
\hline
vs. Legacy-US & TCP & Max & $32.2$ & $80.3$ \\
 & & Median & $12.6$ & $47$ \\
\hline
vs. PUMA & TCP & Max & $82.3$ & $70.2$ \\
 & & Median & $20.1$ & $29.5$ \\
\hline
vs. Legacy-US & UDP & Max & $61.9$ & $132.7$ \\
 & & Median & $7.2$ & $28.9$ \\
\hline
vs. PUMA & UDP & Max & $88.7$ & slightly better than MUSE\\
 & & Median & $76.2$ &  \\
\hline
\end{tabular} 
\label{table:MUSE}
\end{table}

In~\cite{Faridi:2014}, results demonstrate that in short-range WLANs, channel bonding can produce a significant performance enhancement. Performance enhances when external interference signal is low or moderate and dynamic channel bandwidth access mechanism is enabled. Dynamic channel switching between different bandwidths (20/40/80/160 MHz) is based on the result of clear channel assessment (CCA). This type of switching enhances upper layer throughput~\cite{Park:2011}. The MAC protection mechanism and channel bonding can improve cooperatively fairness in applications at upper layer~\cite{Hart:2011}. These works are discussed in Table~\ref{table:13} and Table~\ref{table:14}.

\cite{Oscar:2013} (Table~\ref{table:16}) investigates performance when MU-MIMO is considered with channel bonding. This work discusses two important PHY features -- MU-MIMO and channel bonding. It is observed that these features can improve significantly several network performance metrices such as delay, TCP/UDP throughput etc.

The performance of stream control transmission protocol (SCTP) and IEEE 802.11ac for data link layer and transport layer is evaluated in~\cite{Rung-Shiang:2015} (Table~\ref{table:14}). Results of~\cite{Rung-Shiang:2015} ensure that SCTP produces better transmission performance compared to traditional TCP. Application of SCTP over IEEE 802.11ac creates a faster and more stable wireless network environment. SCTP and IEEE 802.11ac help to realize the vision for broadband transmission in the era of digital convergence. 

Gong \textit{et al.}~\cite{Go:2010} proposed carrier sense multiple access with collision avoidance (CSMA/CA) based MAC protocol using three response mechanisms considering downlink MU-MIMO. A per-station queuing mechanism is designed to tackle the problem of hidden node in wireless network. The proposed scheme reduces end-to-end transmission delay~\cite{Go:2010}. A comparative performance analysis between downlink MU-MIMO and the proposed approach has been carried out. This work and its several issues are discussed in Table~\ref{table:16}.

\noindent{\textbf{Energy efficiency:}} Swetank \textit{et al.}~\cite{Swetank:2015} presented an experimental study of IEEE 802.11n/ac power consumption. Energy efficiency in HT-WLANs is also addressed in~\cite{Prasant:2014} by considering channel bonding and number of spatial streams. Throughput in IEEE 802.11ac is increased by  employing three factors -- larger channel bandwidth, denser modulation and higher number of MIMO spatial streams~\cite{Prasant:2014}. Utilization of larger channel width is a less energy efficient solution. Higher number of spatial streams is more energy efficient~\cite{Prasant:2014}. Thus, TCP/UDP throughput increases with higher MCS values. In~\cite{Swetank:2015}, the impact of different features of IEEE 802.11n/ac (MIMO, frame aggregation, channel bonding etc.) are analyzed by considering both power consumption and throughput. Thus, IEEE 802.11ac increases TCP/UDP throughput and network fairness~\cite{Prasant:2014,Swetank:2015}. A power-efficient protocol exploiting the characteristic of channel bonding is proposed in~\cite{Okhwan:2015}, which helps in reducing end-to-end transmission delay. These works have been highlighted in Table~\ref{table:14}.

To support downlink MU-MIMO at MAC level, TXOP sharing is proposed and it enhances the mandatory TXOP scheme of wireless network standard~\cite{Yazid:2014,Narayan:2015} (Table~\ref{table:15}). Performance evaluation of IEEE 802.11ac is carried out through Markov chain model enabling TXOP sharing technique. Based on the proposed model, throughput is calculated for different access categories. It is observed that enahnced features in IEEE 802.11ac help to increase TCP/UDP throughput for several access categories. 

\subsection{Impact of frame aggregation}

Through analysis and simulations, authors in~\cite{Ong:2011} (Table~\ref{table:14}) compare MAC performance of IEEE 802.11ac with IEEE 802.11n. This comparison considers three frame aggregation mechanisms -- A-MSDU, A-MPDU and hybrid A-MSDU/A-MPDU. Results indicate that IEEE 802.11ac with 80 MHz channel bonding and single spatial stream outperforms IEEE 802.11n with 40 MHz bonding and two spatial streams. Thus, IEEE 802.11ac provides much better throughput. In addition, the hybrid A-MSDU/A-MPDU yields the best upper layer performance for both IEEE 802.11ac and IEEE 802.11n. Therefore, hybrid frame aggregation enhances network capacity.

\subsection{Impact of MU-MIMO, channel bonding and frame aggregation}

Explicit compressed feedback (ECFB) scheme for channel sounding is analyzed in~\cite{Liao:2013}. It proposes an extended RTS/CTS scheme to integrate the ECFB operation. This paper also compares IEEE 802.11ac in saturated conditions. Table~\ref{table:13} mentions this work which shows that several PHY/MAC features can enhance upper layer performance. In~\cite{Eldad:2011} (Table~\ref{table:13}), an overview of IEEE 802.11ac is presented with a focus on MU-MIMO, channel bonding and frame aggregation. Performance of IEEE 802.11ac is analyzed with respect to these features. By enabling these features, it is observed that the overall network capacity enhances in IEEE 802.11ac.

Thus, from existing literature, we may conclude that different enhanced features of PHY/MAC of IEEE 802.11ac can increase upper layer throughput significantly and reduce end-to-end transmission delay. Several comparative studies related to IEEE 802.11ac are presented in Table~\ref{table:13} -- Table~\ref{table:16}.

\section{Future Research Directions on the Design of IEEE 802.11n based Networks}
\label{sec:n}

As discussed, IEEE 802.11n introduces some novel PHY/MAC features and these changes have an impact over TCP/UDP throughput. However, weaknesses as reported in the Tables  in Appendix indicate that we need to do further research to achieve high goodput at application level performance which is based on HT-WLANs. Open research challenges in this direction are summarized below. 
 
\subsection{Possible Extensions of Existing Works}
Further directions in research on IEEE 802.11n are discussed in the following.

\subsubsection{Design based Research Directions}

In this subsection, we discuss design challenges of existing works.

\noindent{\textbf{Design of scheduler for frame aggregation:}}

A-MSDU and A-MPDU sizes affect the average network throughput and average transmission delay for various TCP congestion control methods~\cite{Xia:2009}. As the aggregation size increases, the number of subframes within an aggregated frame also increases~\cite{Sandip:2014}. It reduces MAC layer overhead and increases MAC layer throughput. However, throughput becomes saturated beyond a certain size of aggregation. It is because loss of one aggregated frame causes a loss of a huge number of subframes. Performance of video streaming using frame aggregation is evaluated in~\cite{Liang:2014}. 

End-to-end transmission delay decreases as the aggregation size increases since more frames can be delivered within a given period of time in form of subframes in an aggregated frame. It also becomes saturated beyond a size of frame aggregation. Results in~\cite{Mustafa:2012} show that TCP-NewReno and TCP-SACK outperform other TCP variants using different aggregation sizes. Therefore, an appropriate size of frame aggregation should be used to achieve the maximum TCP throughput under different congestion control mechanisms. An appropriate mixture of A-MSDU and A-MPDU needs to be considered to reduce MAC overhead and increase data rate~\cite{Daniel:2012}. An appropriate scheduler combining frame aggregation and BACK considering all existing studies needs to be designed.

\subsubsection{Evaluation based Research Directions}

Challenges regarding evaluation of proposed mechanisms in the existing literature are summarized in the following.

\noindent{\textbf{Performance evaluation of TCP congestion control mechanisms:}}

Enhancement of MAC layer throughput may not always enhance transport layer throughput. For example, TCP traffic uses congestion control mechanisms. Different congestion control mechanisms achieve different throughputs for the same traffic load, frame aggregation schemes~\cite{Gong:2012} and PHY specifications (MIMO, MCS, number of spatial streams)~\cite{Xia:2009}. Performances of different TCP congestion control schemes can be evaluated using PHY enhancements like channel bonding, OFDM-MIMO, different MCS values and short guard interval. There are many frame aggregation algorithms in IEEE 802.11n as discussed in Tables~\ref{table:4} -- Table~\ref{table:12}. Some of these algorithms use different traffic categories. All these algorithms can be evaluated under various TCP congestion control mechanisms to get more insight about the impact of frame aggregation over TCP/UDP goodput.

\noindent{\textbf{Evaluation of the effect of frame aggregation over TCP/UDP performance in AP association in IEEE 802.11n:}}

For heterogeneous client stations, an AP association mechanism is designed for IEEE 802.11n in~\cite{Gong:2012}. It utilizes full benefits of frame aggregation but MAC level efficiency is examined in this work. To understand the performance of upper layer protocols, TCP/UDP throughput has to be evaluated for this association scheme.

\noindent{\textbf{Application of block acknowledgement in TCP/UDP traffic evaluation:}}

Tables~\ref{table:4} to~\ref{table:12} highlight research works which consider block acknowledgement mechanisms. For example, a compressed block acknowledgement scheme is designed in \cite{Tassiulas:2008}. This new MAC feature can be examined in TCP and UDP traffic along with frame aggregation, QoS and power save mechanisms.

\noindent{\textbf{TCP/IP capability investigation in Vehicular Internet Environment:}}

TCP/IP performance of IEEE 802.11n is found to be optimal in low data rate under Vehicular Internet Environment~\cite{Fang:2009}. All new features of IEEE 802.11n have to be incorporated in this performance evaluation. PHY and MAC enhancements of IEEE 802.11n have to be considered to find TCP/IP capabilities in vehicular Internet environment. Note that vehicular mobility imposes a special type of mobility model that may impact underlying link adaptation algorithms which can further be extended to application layer performance.

\subsection{Open Challenges}
There are many challenges associated with IEEE 802.11n, which have to be resolved to practically obtain high throughput. These challenges are discussed in the following. 

\subsubsection{Design based Research Directions}

Design related challenges are presented in the following.

\noindent{\textbf{Design of effective channel bonding mechanisms:}}

An intelligent channel bonding method is discussed in~\cite{Deek:2014}. However, this work and similar other works do not consider impact of channel bonding on application performance. The challenge is to design an efficient and intelligent channel bonding algorithm to improve the performance of transport layer protocols.

\subsubsection{Evaluation based Research Directions}

Limitations in research to evaluate interaction of IEEE 802.11n and transport layer protocols are stated in the following.

\noindent{\textbf{Exploit the interaction between IEEE 802.11n and transport layer protocols:}}

Protocols like TCP is negatively impacted by channel bonding at high MCS. Therefore, the current implementation of IEEE 802.11n is able to provide physical data rates less than 250 Mbps. In~\cite{Sandip:2014}, it was observed that physical and MAC enhancements in IEEE 802.11n do not translate fully into the performance of transport layer protocols like TCP. This study provides a challenge for researchers to exploit interactions between IEEE 802.11n and transport layer protocols to exploit the full capacity of physical high data rate communication and achieve higher throughput in WLANs. Hence, an effective co-ordination mechanism has to be designed between MAC protocol and transport protocols.

\noindent{\textbf{Performance evaluation of OFDM-MIMO:}}

OFDM-MIMO is an important enhancement of PHY layer. OFDM-MIMO enhances the range of data transmission and TCP throughput~\cite{Richard:2006}. The performance of OFDM-MIMO can be examined using channel bonding, short guard interval and frame aggregation. So, TCP throughput can be evaluated with OFDM-MIMO along with MAC layer enhancements.

\section{Future Research Directions for IEEE 802.11ac Networks}
\label{sec:ac}

Use of a wider channel increases the chance of suffering from external interference. However, faster transmission compensates the negative impact of the external interference. In this way, all new features introduced in IEEE 802.11ac have an impact on system performance. However, there are many challenges to obtain theoretically achievable high throughput practically in IEEE 802.11ac. Some of these challenges are summarized in the following subsections.

\subsection{Possible Extensions of Existing Works}
Future works as recorded in the existing literature for IEEE 802.11ac are discussed in the following.

\subsubsection{Design based Research Directions}

In IEEE 802.11ac, design issues raised in the existing literature are mentioned in the following.

\noindent{\textbf{Effective design of Explicit Compressed Feedback (ECFB) channel sounding policy:}}

The ECFB discussed in~\cite{Liao:2013} is a channel sounding technique that should be designed in a novel way to reduce overhead in a non-saturated condition. To make a request of on-demand CSI to some specific stations is the easiest option to overcome these overheads. This type of request can be made only when the sender has data directed to the receiver. Effective design of ECFB channel sounding policy is also a challenge in IEEE 802.11ac. That may help to achieve high throughput.

\subsubsection{Evaluation based Research Directions}

Challenges in evaluation of existing mechanisms in IEEE 802.11ac are summarized as follows.

\noindent{\textbf{Estimation of the overhead loss of IA and coordinated multi-point (CoMP) under non-stationary conditions:}}

Gains of IA and CoMP over TDMA-based MIMO as discussed in~\cite{Per:2014} are $30\%$ and $70\%$ respectively under stationary conditions. In dense deployment, the frequency domain granularity of the feedback can be reduced to about every 8th subcarrier ($5$ MHz), without sacrificing performance. One future work could be to find the update rate needed to estimate the loss of IA and CoMP under non-stationary conditions.

\noindent{\textbf{Performance evaluation of Multi-packet reception (MPR) with Multi-user RTS/CTS (MU-RTS/CTS):}}

Simultaneous transmissions to multiple users help to increase the overall network throughput~\cite{Esslaoui:2012}.
A MU-RTS/CTS mechanism discussed in~\cite{Liao:2013}, allows a station to transmit frames to more than one station simultaneously. However, MPR can reduce chances of collisions and thus, it helps to improve the overall system performance. MPR requires a synchronization among distributed wireless stations, which applies MU-RTS/CTS handshaking. It is a better approach than simple Multi User Basic (MU-Basic). So, MU-Basic should be extended to allow MPR. The performance of MPR along with MU-RTS/CTS can be evaluated in IEEE 802.11ac and it is a challenge to achieve higher upper layer throughput through this mechanism.

\noindent{\textbf{Evaluation and modification of CSMA/CA based downlink MU-MIMO (DL MU-MIMO):}}

Evaluation of CSMA/CA based DL MU-MIMO protocol~\cite{Go:2010} using channel bonding, frame aggregation and short guard interval is a future work. Modification of this mechanism considering all enhanced features of IEEE 802.11ac is a challenge to achieve very high throughput in upper layer.

\subsection{Open Challenges}
There are many open issues in IEEE 802.11ac, which are mentioned in the following.

\subsubsection{Design based Research Directions}

Design related challenges regarding IEEE 802.11ac are discussed in the following.

\noindent{\textbf{Impact of reverse direction feature and MU-MIMO:}}

Reverse direction method and MU-MIMO can also be applied in the performance evaluation of TCP and UDP to observe their effects on upper layer. Existing studies do not consider about the impact of reverse direction over application performance. 

\noindent{\textbf{Application of Beamforming technology:}}

Beamforming technology is a PHY layer enhancement in IEEE 802.11ac~\cite{Femi:2013}. Beamforming technology could be modified to get better performance in HT-WLANs and thus, to achieve very high throughput at upper layer.

\noindent{\textbf{Consideration of MAC and the network routing protocols together:}}

Hidden nodes reduce the performance of other nodes and decrease network throughput. Thus, it should be considered in wireless multi-hop mesh network. MAC and network routing protocols are needed to be considered together since there are multiple stations as destinations in single MU-MIMO data transmission. This improvement is a challenge in IEEE 802.11ac.

\noindent{\textbf{Design of an adaptive scheduling algorithm considering different enhanced parameters of IEEE 802.11ac:}}

The number of wireless stations and the size of the queue in a station have a significant impact on the performance of a system in a non-saturated condition~\cite{Zhiqun:2015}. When the number of stations is very high, heterogeneous types of destinations in packets make it difficult to achieve full advantage of frame aggregation in MU-MIMO based IEEE 802.11ac WLANs. Existing schedulers do not consider all PHY/MAC features of IEEE 802.11ac~\cite{Aajami:2015},~\cite{Esslaoui:2012},~\cite{Juan:2015}. As a consequence, the challenge is to design an effective scheduler to perform frame aggregation to achieve high throughput in upper layer. Designing an effective multi-level frame aggregation for different access categories using the concept of dynamic channel management is a great challenge in IEEE 802.11ac. This approach can play an important role to increase TCP/UDP throughput and thus, it can meet high system throughput.

An adaptive scheduling algorithm which can jointly consider different novel parameters of IEEE 802.11ac may play an important role to improve system performance. There are several factors which influence the system throughput such as spatial stream per frame allocation, size of aggregated frames (A-MSDU and A-MPDU), number of nodes per antennas, channel bandwidth, queuing state information and channel interference conditions. In this regard, it is also needed to consider some points such as minimization of frequency of channel sounding, maximization of system throughput and minimization of unfairness to active stations. Designing an adaptive scheduling algorithm considering all these parameters is a great challenge in very high throughput WLANs.

\subsubsection{Evaluation based Research Directions}

In evaluation, some works need to be followed to address open challenges in IEEE 802.11ac.

Open challenges in evaluation of mechanisms in IEEE 802.11ac are summarized as follows.

\noindent{\textbf{Determination of appropriate combination of configuration parameters:}}

From Table~\ref{table:13} -- Table~\ref{table:16}, we may conclude that the use of channel bonding in combination with other mechanisms in IEEE 802.11ac such as multi-user beamforming and multi-level frame aggregation is still an open challenge. In this respect, several new trade-offs have to be considered. For instance, for a given transmission power, the use of wider channels implies that the signal to noise ratio per subcarrier will be lower. This situation may require the selection of more robust modulation schemes and coding rates. Otherwise, the gain in throughput may be lost. When multi-user beamforming is considered, the use of wider channels requires a larger amount of CSI~\cite{Hyunsub:2014} which is fed back to the AP from stations. Hence, network overhead is increased. Finally, frame aggregation is needed to reduce high frame and protocol overheads in MAC layer. Also, the gain in throughput provided by channel bonding can be achieved. However, it may not be a good solution if secondary channels have a high alternating rate between busy and free states. In this case, transmissions containing multiple frames have a higher chance of being corrupt by external interference. Aforementioned trade-offs need to be carefully studied in order to determine an appropriate configuration parameters in each specific scenario such that very high throughput can be achieved in upper layer.

\noindent{\textbf{Effect of dynamic bandwidth management:}}

Dynamic bandwidth management is a new PHY feature introduced in IEEE 802.11ac~\cite{Park:2011}. This feature was not included in IEEE 802.11n. The aim of this mechanism is to produce better throughput than IEEE 802.11n. Dynamic bandwidth management is introduced to manage the set of primary channels and secondary channels in wide bandwidth conditions. Hence, this feature should be considered for performance evaluation of IEEE 802.11ac. Further, the impact of this mechanism over upper layer performance has to be evaluated.

\section{General Research Challenges: Directions for Future HT-WLAN Developments}
\label{sec:nac}
In this section, we summarize future research directions which are common to both high throughput wireless network extensions -- IEEE 802.11n and IEEE 802.11ac. 

\subsection{Possible Extensions of Existing Works}

We discuss possible extensions of existing research works in the following.

\subsubsection{Design based Research Directions}

For both IEEE 802.11n and IEEE 802.11ac, design related challenges of existing works are discussed in the following.

\noindent{\textbf{Design of MU-MIMO antenna:}}

MU-MIMO was introduced in IEEE 802.11ac and it can increase throughput of wireless networks significantly~\cite{Zhu:2011,Boris:2012,Esslaoui:2012,Panajotovic:2013,Panajotovic:2015}. MU-MIMO can also be incorporated into IEEE 802.11n. Hence, there is a design issue of MU-MIMO antenna in both IEEE 802.11n and IEEE 802.11ac.

In~\cite{Keke:2014}, Tomlinson-Harashima Precoding (THP) processing technique is employed to design MU-MIMO systems. In this work, it is shown that the performance of THP depends heavily on the sequence of precoded symbols. The optimal ordering algorithm is impractical for MU-MIMO with multiple receive antennas as users are geographically distributed. Proposed THP-based algorithms in~\cite{Keke:2014} apply degrees of freedom to transmit data by constructing a set of branches, that lead to an extra transmit diversity gain. Javier \textit{et al.}~\cite{Javier:2013} addressed the problem of design of precoder in a MIMO broadcast environment with battery-powered terminals. In this work, power consumption model of radio frequency (RF) and decoding stages are included in the design of the proposed scheme. The sum-rate maximization is also considered in this mechanism as an optimization policy.

Authors in~\cite{Mo:2016} proposed a secure transmission scheme by applying various beamforming matrices to data signals and pilot signals. In this work, the beamforming matrix is generated in a way that legitimate users can derive the channel matrix. The generated matrix is experienced by data signals on the basis of pilot signals. The proposed beamforming scheme in~\cite{Mo:2016} is formalized as an optimization problem that maximizes the minimum signal-to-interference-plus-noise ratio (SINR). In~\cite{Yongjiu:2015}, frequency selectivity of current wireless channel bandwidths is considered to optimize MU-MIMO. Further, the occupied channel bandwidth is divided into equal sub-channels according to the level of selection of frequency. Each sub-channel is allocated as per the largest bandwidth.

Considering all aforementioned works, we can see that proposed schemes and methodologies may be employed in designing MU-MIMO spatial streams in HT-WLANs. These developments could help to enhance transport/application layer throughput as MU-MIMO increases the overall system performance in IEEE 802.11n/ac.
\subsubsection{Evaluation based Research Directions}

Evaluation related shortcomings of researches in IEEE 802.11n/ac are stated in the following.

\noindent{\textbf{Impact of channel bonding over upper layer protocols using QoS and power save mode:}}

An evaluation of the effect of channel bonding over upper layer protocols using QoS and power save mode (U-APSD, S-APSD, IEEE 802.11 power save mode) can be undertaken. QoS and power save mode are important factors which can be incorporated with enhancements of PHY and MAC in IEEE 802.11n. Thus, these factors may lead to increase in the goodput~\cite{Janis:2012}. A basic problem in IEEE 802.11 is to find out the cause of packet loss accurately. It is becoming increasingly important since wireless data rate is scaling to Gbps. In~\cite{Anwar:2014}, a loss differentiation framework, Burst Loss Monitor (BLMon), has been proposed. BLMon can differentiate losses accurately with minimal overhead. It applies loss patterns in aggregate frames. Aggregate frames retry to achieve loss differentiation accurately with low overhead. This model can be used with channel bonding and power save mode. As a result, we can find the effect of this model over transport layer protocols in both IEEE 802.11n and IEEE 802.11ac.

\subsection{Open Challenges}

In this section, we discuss open challenges of both IEEE 802.11n and IEEE 802.11ac related works with respect to both design and evaluation points of view.

\subsubsection{Design based Research Directions}

Design related issues of IEEE 802.11n/ac are summarized in the following.

\noindent{\textbf{Design of link adaptation mechanism:}}

Link adaptation is a very important issue in WLAN. There are many rate adaptation mechanisms in IEEE 802.11 WLANs~\cite{Chunyi:2015,Anwar:2012,Saif:2013,Leo:2002,Hoss:2009,Madwifi,Samplerate,
sample:2015}. But, none of them are able to meet theoretically achievable throughputs in practical scenarios in IEEE 802.11n/ac. Moreover, there is no such rate adaptation algorithm which can consider all enhanced features introduced in IEEE 802.11n/ac. A performance analysis of ARAMIS (Agile Rate Adaptation for MIMO Systems) discussed in~\cite{Deek:2013} can be evaluated using TCP and UDP traffic. To find out the impact of Minstrel rate adaptation algorithm over transport layer protocols in IEEE 802.11n is also a challenge. A combination of intelligent channel bonding and different rate adaption algorithms can be applied to find their impact over upper layer protocols in IEEE 802.11n. IEEE 802.11ac should also adapt to dynamic link condition to achieve the maximum physical data rate~\cite{Khavasi:2014},~\cite{Aajami:2015}. 

In one of our works, we designed a cross-layer link adaptation mechanism~\cite{karmakar2015dynamic} which considers channel bonding, SGI, frame aggregation and various MCS levels in IEEE 802.11n. In the analysis section, we have shown that performance of a wireless system deteriorates as the number of stations increases or the signal strength of the channel decreases. In our another work~\cite{Karmakar2016bandit}, we developed an intelligent distributed link adaptation scheme in IEEE 802.11ac considering channel bonding, frame aggregation, SGI and MCS levels. In this work, it is also observed that low signal quality and high number of stations affect the overall network performance negatively. As a result, proper tuning of PHY/MAC features of HT-WLANs is necessary according to the channel condition for high quality link adaptation.

Hence, designing an efficient link adaptation mechanism is very much needed to gain very high throughput in dynamic channel environment. Abedi \textit{et al.}~\cite{Abedi:2016} proposed a scheme for estimation of frame error rate (FER) of one PHY/MAC configuration set from FER of another set. Thus, to design an optimization algorithm that can find optimal link configuration set, it is not needed to find FERs of all configurations. Using the methodology proposed in~\cite{Abedi:2016}, a subset of all configurations in HT-WLANs can be used to find performance of all such sets. As a result, an efficient link adaptation algorithm can be developed from a small PHY/MAC configuration set.

MUSE rate adaptation scheme considers MU-MIMO and channel bonding features of IEEE 802.11ac~\cite{Sur:2016}. The compressed beamforming feedback used by MUSE can also be engaged to apply dynamic channel management mechanism. Further, in low congested scenario, SGI can be activated. ESNR proposed in~\cite{Halperin:2010}, selects rate and delivers packets on the basis of CSI. But, it applies only MIMO in IEEE 802.11n. As a consequence, none of MUSE and ESNR fully exploit HT-WLANs. So, a learner can be designed for both MUSE and ESNR, using a machine learning technology such as multi-armed bandit, reinforcement learning etc. Depending on beamforming feedback, this learner will tune data rate using MU-MIMO, channel bonding and SGI. Also, it will be able to perform dynamic channel management. So, the challenge is to develop an effective rate adaptation algorithm considering new features of IEEE 802.11n/ac to achieve very high TCP/UDP throughput.

\noindent{\textbf{Design of efficient power management mechanism:}}

Through experiments, Swetank \textit{et al.}~\cite{Swetank:2015} investigated the impact of different features of IEEE 802.11n/ac over throughput, energy consumption and their trade-offs. Energy efficiency using enhanced PHY features and the study of interference characterization are important issues in IEEE 802.11ac~\cite{Prasant:2014}. In this work, through IEEE 802.11ac testbed experiments, it is shown that utilization of larger channel width is less energy efficient for higher power consumption during idle listening mode. In~\cite{Okhwan:2015}, it is experimentally verified that wider bandwidth consumes more energy, whereas power consumed during idle or receive (RX) state depends considerably on channel bandwidth which is quite comparable to transmit (TX) state power. By these observations, a power save operation is proposed in~\cite{Okhwan:2015}.

Ucar \textit{et al.}~\cite{Inaki:2016} addressed the relationship between energy efficiency and throughput performance in multi-rate IEEE 802.11 network. In this work, it is experimentally shown that a small degradation in throughput may result in energy efficiency. Trade-off between energy efficiency and throughput arises due to the power consumption of wireless cards. These studies imply that rate adaptation mechanism can be developed by taking power consumption of wireless interface into account. One may attempt to design a distributed energy-efficient model for HT-WLANs where each wireless station is able to select its data rate. In such efforts, the effect of different PHY/MAC configurations over power consumption should be examined. On its basis, new link adaptation schemes can be developed for HT-WLANs. Then, the impact of such rate adaptation schemes over upper layer protocols can be evaluated.

None of these works consider all new features of IEEE 802.11ac and thus, can not evaluate the effect of different combinations of PHY and MAC features over power consumption. Energy consumption is a serious issue in smartphones. Hence, power efficient protocols should be designed considering all PHY and MAC features of IEEE 802.11n/ac. Then, it will be useful for next generation mobile phones. In addition to very high throughput, energy efficiency will improve the network performance of upper layer in such devices.

\noindent{\textbf{Design of multilevel frame aggregation scheduler with block acknowledgement:}}

Frame aggregation and block acknowledgement help to reduce MAC overhead. Their collective contributions can further enhance MAC efficiency. Several existing literature discussed in Table~\ref{table:4} -- Table~\ref{table:16} consider these two new MAC features. In these tables, we have discussed the strengths and weaknesses of works related to frame aggregation and BACK and have also explained impacts of these works over transport/application layer protocols. Frame aggregation mechanism in IEEE 802.11n is discussed in~\cite{Kolap:2012}. This work shows how a station can combine several MSDUs for transmitting data to multiple destinations. Then, the aggregated AMSDU is transmitted to AP. Thereafter, the AP packs multiple MSDUs (with the same destination address) from different source addresses and sends them to a single destination. Thus, an A-MSDU frame is not transmitted to multiple receivers. All MPDUs in an A-MPDU carry the same TID to work with BACK mechanism efficiently, in which A-MPDU is different for each user.

So, A-MSDU and A-MPDU can work cooperatively to enhance efficiency of frame aggregation technique. It is a great challenge to combine both A-MSDU and A-MPDU to create a multilevel scheduler to take full advantage of frame aggregation. Further, block acknowledgement can also be incorporated into multilevel frame aggregation. Hence, the open challenge is to design an efficient multilevel frame aggregation mechanism combined with block acknowledgement scheme.

\subsubsection{Evaluation based Research Directions}

Challenges regarding evaluation of IEEE 802.11n/ac related mechanisms are stated as follows.

\noindent{\textbf{Effect of multicast transmission:}}

Multicast transmission can improve throughput by transmitting data to multiple stations simultaneously. In wireless environment, multicast transmission related issues are investigated in~\cite{Dolnak:2015}. These issues have raised some points such as dynamic routing protocols which are examples of multicast communications in wireless networks. Nowadays, effective network routing is not possible without multicast transmission. However, we need specific optimization in multicast transmission in wireless environment and various multicast issues need to be explored in various network layers. Thus, an optimization of multicast transmission is required for improving dynamic routing mechanisms which further enhance performance of network layer. As a result, the application throughput can be improved. In HT-WLANs, analyzing the effect of multicast transmission over upper layer protocols using channel bonding, frame aggregation and BACK is also an open challenge in IEEE 802.11n/ac.

\noindent{\textbf{Effect of mobility:}}

Mobility can produce a great impact over performance of HT-WLANs. It may degrade the overall network performance due to dynamic positions of wireless stations~\cite{Janis:2012}. Chen \textit{et al.}~\cite{chenMobility:2012} carried out some experiments to show that rapid fluctuation of channel conditions and channel asymmetry are two salient features of wireless channel in mobile environment. Thus, it becomes critical to have an efficient rate adaptation scheme in IEEE 802.11, which can handle these issues. A rate adaptation mechanism called Rate Adaptation in Mobile environments (RAM) was designed and implemented for IEEE 802.11 legacy standards in MadWifi device driver~\cite{chenMobility:2012}. This type of scheme can also be developed for HT-WLANs by taking mobility into account.

Different enhanced features of PHY/MAC in HT-WLANs highly depend on various parameters of wireless channel such as signal strength, interference etc. Values of these parameters influence the performance of enhanced features of HT-WLANs. As the channel condition changes rapidly in a mobile environment depending on the position of wireless station, the overall performance of HT-WLANs depends on mobility. Effect of mobility on intelligent channel bonding mechanism as proposed in~\cite{Deek:2014} can also be examined using TCP/UDP traffic along with frame aggregation and BACK. Effects of mobile environment also need to be considered while developing any new mechanism/protocol in IEEE 802.11n/ac high throughput WLANs.

\noindent{\textbf{Channel Access Fairness:}}

Enhanced features of PHY and MAC of HT-WLANs have many internal trade-offs in different channel conditions~\cite{karmakar2015dynamic}. Channel bonding may provide unfairness because neighboring APs choose bonding levels independently. In one of our works, we have investigated the effect of channel bonding on throughput fairness in a testbed experiment~\cite{Karmakar2016fairness}. We grouped several STAs on the basis of bands they used. Table~\ref{tab:t1} describes scenarios used in the experiment, whereas results are shown in Figure~\ref{fig:g1} and Figure~\ref{fig:g2}. These figures indicate that there is a high impact of co-channel interference over Jain's Fairness Index (JFI). Also, dynamic bandwidth channel management suffers from interference (Figure~\ref{fig:g1}). For overlapping channels (S9-S12), throughput deviation is very high and it is due to the inter-channel interference (Figure~\ref{fig:g2}).

\begin{table}[!ht]
\tiny
 \caption{Testbed Configuration Scenarios (DBCA: Dynamic Bandwidth Channel Access, U: Undefined, NO: Non Overlapping Channels, O: Overlapping Channels)~\cite{Karmakar2016fairness}}
 \label{tab:t1}
\centering
 \begin{tabular}{l||l|l}
  \hline
  \textbf{Seq} & \textbf{Scenario} & \textbf{Type}\\
  \hline \hline
  S1 & All: DBCA & U\\
  S2 & All: $20$ MHz & NO\\
  S3 & All: $40$ MHz & NO\\
  S4 & All: $80$ MHz & NO\\
  S5 & R1,R4,R5: $20$ MHz; R2,R3,R6: $40$ MHz & NO \\
  S6 & R1,R4,R5: $40$ MHz; R2,R3,R6: $80$ MHz & NO \\
  S7 & R1,R4,R5: $80$ MHz; R2,R3,R6: $40$ MHz & NO \\
  S8 & R1,R4,R5: $20$ MHz; R2,R3,R6: $80$ MHz & NO \\
  S9 & R1,R4,R5: $20$ MHz; R2,R3,R6: $40$ MHz & O ($20$) \\
  S10 & R1,R4,R5: $40$ MHz; R2,R3,R6: $80$ MHz & O ($20$)  \\
  S11 & R1,R4,R5: $40$ MHz; R2,R3,R6: $80$ MHz & O ($40$) \\
  S12 & R1,R4,R5: $20$ MHz; R2,R3,R6: $80$ MHz & O ($20$) \\
  \hline 
 \end{tabular}
\end{table}

\begin{figure}[!ht]
\begin{minipage}[t]{0.45\linewidth}
\hspace{5mm}
   \includegraphics[scale = 0.312]{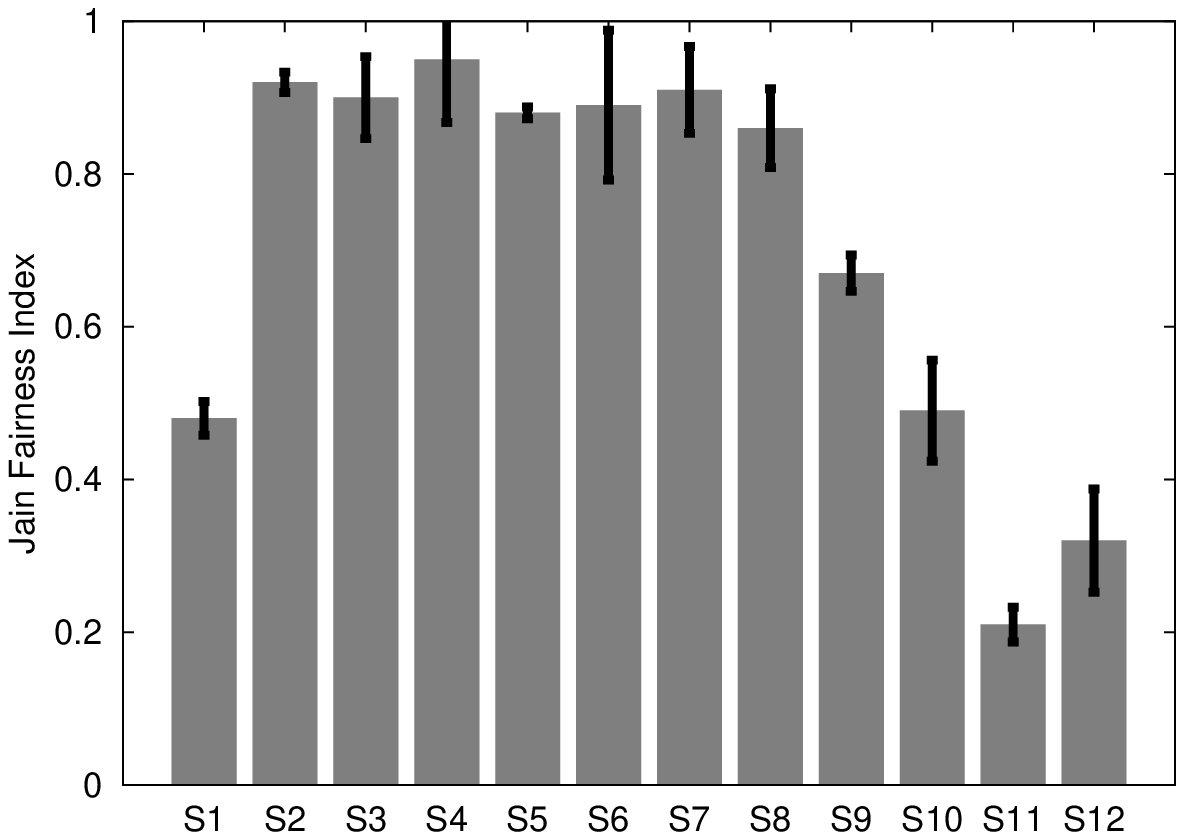}
    \caption{Jain Fairness Index (JFI) for the $12$ scenarios: S9-S12 have minimum fairness index indicating that co-channel interference is a serious issue~\cite{Karmakar2016fairness}}
    \label{fig:g1}
\end{minipage}%
    \hfill%
\begin{minipage}[t]{0.45\linewidth}
\hspace{4mm}
   \includegraphics[scale = 0.312]{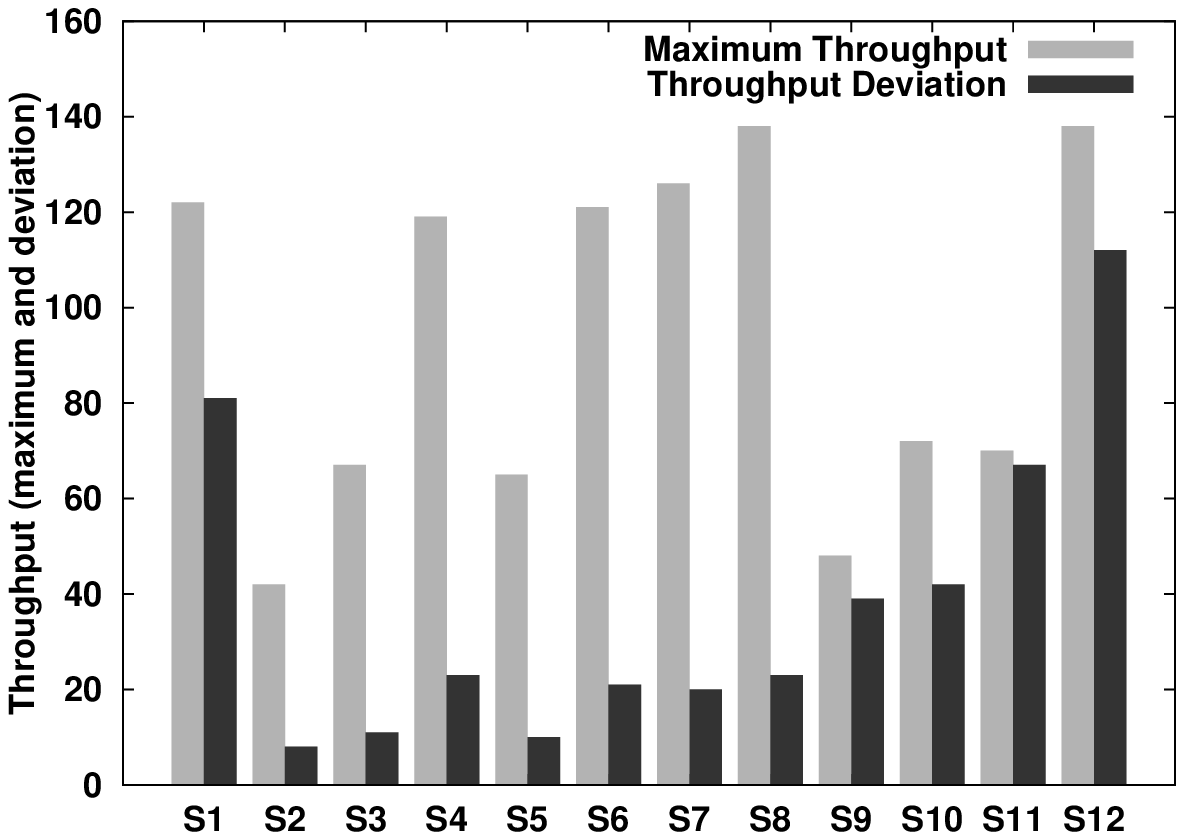}
    \caption{Maximum STA throughput and throughput deviation for the $12$ scenarios: S9-S12 have maximum throughput deviation~\cite{Karmakar2016fairness}}
    \label{fig:g2}
\end{minipage} 
\end{figure}

Therefore, new MAC layer mechanism should be designed to enhance channel access fairness as we proposed in~\cite{Karmakar2016fairness}. In this work, we also evaluated the fairness of TCP/UDP throughput of the proposed approach. Design of a fairness maximization algorithm for upper layer considering all enhanced features of HT-WLANs is a challenging issue.

\section{Conclusion}
\label{sec:conc}

This survey provides a deep study of fundamental concepts and mechanisms of high throughput WLANs with their impacts on application performance. We have considered IEEE 802.11n amendment which provides high throughput by extending PHY and MAC features. We have also considered IEEE 802.11ac which provides very high throughput and this specification is an extension of the latest IEEE 802.11n. Main innovative concepts which have been introduced in MAC layer of HT-WLANs are frame aggregation and BACK mechanism. High throughput standards can achieve a maximum data rate of $7$ Gbps (IEEE 802.11ac) with advanced signal processing mechanisms such as MIMO spatial multiplexing, transmit beamforming etc. These techniques enhance the range of WLANs significantly. Additionally, system throughput and reliability also increase. Further, HT-WLANs increase the channel capacity using wide frequency bands (maximum of 160 MHz in IEEE 802.11ac) with MIMO-OFDM technique. Channel bonding and OFDM-MIMO are new enhanced features in HT-WLANs. The MIMO technology improves performance since it relies on antenna diversity and spatial multiplexing. This type of multiplexing allows a wireless station to transmit and receive from multiple spatial channels at once. IEEE 802.11ac can support a maximum of $8\times 8$ MIMO streams and thus, it can provide better performance than legacy standards. In IEEE 802.11ac , MAC layer supports frame aggregation mechanism with larger frame size as well as a new TXOP sharing technique. This standard can also use downlink MU-MIMO technology (up to four clients). 

We have surveyed the impact of enhancements of PHY and MAC layer of IEEE 802.11n and IEEE 802.11ac on transport/application layer performance. We have reviewed different enhanced PHY and MAC schemes which have some effect on the performance of upper layers. In this survey, we have shown that few research works have considered all new features of PHY and MAC layers of high throughput WLANs. Some of them have discussed about TCP/UDP throughput. Hence, the evaluation of performance of upper layers using all enhanced parameters of PHY/MAC is still an open research area in HT-WLANs. Link adaptation mechanism is also an important issue to adjust data rate dynamically depending on the channel condition. It is a challenge to develop an efficient link adaptation mechanism in HT-WLANs to achieve theoretically achievable throughput in practical scenarios. To compute the impact of these PHY/MAC enhancements on TCP/UDP performance in different network scenarios is a future research area in HT-WLANs. 

The physical layer capacity of wireless technology has been boosted up significantly by recent developments. In spite of this enhancement, the improvement of performance in upper layers remains still a big challenge in next generation WLANs.


\bibliographystyle{IEEEtran}
\bibliography{references1}

\onecolumn
\appendix[Summary of Exiting Works on the Impact of High Throughput IEEE 802.11 Extensions]\label{appA}


\begin{table*}[h]
\caption{Consideration of one enhanced feature of HT-WLANs}
\centering
\begin{tabular}{|p{1cm}|p{1.6cm}|l|l|l|l|l|p{2cm}|p{1cm}|p{2.7cm}|p{2.7cm}|}
\hline
\textbf{Item} & \textbf{Objective} & \multicolumn{5}{c|}{\textbf{Trade-off}} & \textbf{Impact on upper layer protocols} & \textbf{Category} & \textbf{Strength} & \textbf{Weakness} \\
\cline{3-7} 
 & & \textbf{MIMO} & \textbf{CB} & \textbf{SGI} & \textbf{FA} & \textbf{BA} & & & & \\
\hline

Liang \textit{et al.} \cite{Liang:2014} & Design of rate adaptation algorithm & Yes & No & No & No & No & Finds optimal rate on per packet basis and thus upper layer throughput can be improved & IEEE 802.11n & 

1) MU-MIMO,
2) Adaptive bit rate on per-packet basis, 
3) No extra control frame
& 
1) Design and placing of MU-MIMO antennas are not discussed, 
2) MAC layer is not involved
\\
\hline

Adamu \textit{et al.} \cite{Adamu:2015} & Design of two-level frame aggregation & No & No & No & Yes & No & Increases throughput by enhancing channel utilization & IEEE 802.11n & 

1) A-MSDU with minimal headers overhead,
2) Both MSDU and MPDU levels are considered, 
3) Maximizes utilization of PHY efficiency at MAC level
& 
1) Does not specify implementation details, 
2) Does not consider PHY enhanced feature to maximize PHY efficiency
\\
\hline

Gong \textit{et al.} \cite{Gong:2012} & AP association mechanism for IEEE 802.11n using heterogeneous clients & No & No & No & Yes & No & Improves the overall network architecture and AP association delay & IEEE 802.11n & 

 1) Formulation of optimization problem using Markov model, 
 2) Heuristic AP association, 
 3) Coexistence of legacy standards
 & 
 1) Different channel conditions are not included,
 2) Noise and interference levels are not mentioned,
 3) End-to-end performance issue is not specified
 \\
\hline

Gong \textit{et al.} \cite{Gong:2013} & Routing and frame aggregation strategy & No & No & No & Yes & No & Reduces routing delay & IEEE 802.11n & 

 1) Distributed algorithm,
 2) Frame aggregation from traffic flow
 & 
 1) Does not provide analysis of routing efficiency,
 2) Time bound analysis of two algorithms is not given,
 3) Does not consider channel error condition
 \\
\hline

Anwar \textit{et al.} \cite{Anwar:2013} & A-MSDU Frame Aggregation Scheme & No & No & No & Yes & No & Helps to reduce end-to-end delay & IEEE 802.11n & 

 1) Sub-frame integrity check, 
 2) Retransmission at MSDU level without altering MAC frame header,
 3) Adding control fields to subframe header
 & 
 1) BA is not applicable at A-MSDU,
 2) New field enhances MAC overhead
 \\
\hline

Sandip \textit{et al.} \cite{Sandip:2015} & Controlling unfairness due to physical layer capture and channel bonding in IEEE 802.11n+s & No & Yes & No & No & No & Adaptive channel bonding improves end-to-end flow performance and network capacity & IEEE 802.11n & 

 1) Consideration of physical layer capture,
 2) Adaptive bonding opportunity and channel reservation scheme 
 & 
 Performance evaluation is carried in only IEEE 802.11n 
 \\
\hline

Navid \textit{et al.} \cite{Navid:2011} & Advanced aggregation techniques to improve fiber-wireless network throughput & No & No & No & Yes & No & New network architecture reduces delay for video, voice and data traffic & IEEE 802.11n & 

 1) A novel fiber-wireless backhaul network architecture,
 2) Advanced hierarchical aggregation,
 3) Different traffic categories
 & 
 1) Does not provide implementation details,
 2) Analysis of end-to-end flow is not given
 \\
\hline

Vanaja \textit{et al.} \cite{Vanaja:2013} & Bi-scheduler algorithm for frame aggregation & No & No & No & Yes & No & Reduces end-to-end delay of packet transmission & IEEE 802.11n & 

 Segregates frames based on their access categories &

 Overhead increases due to two schedulers 
 \\
\hline
\end{tabular} 
\label{table:4}
\end{table*}

\begin{table*}
\caption{Consideration of one enhanced feature of HT-WLANs (continue...)}
\centering
\begin{tabular}{|p{1cm}|p{1.6cm}|l|l|l|l|l|p{2cm}|p{1cm}|p{2.7cm}|p{2.7cm}|}
\hline
\textbf{Item} & \textbf{Objective} & \multicolumn{5}{c|}{\textbf{Trade-off}} & \textbf{Impact on upper layer protocols} & \textbf{Category} & \textbf{Strength} & \textbf{Weakness} \\
\cline{3-7} 
 & & \textbf{MIMO} & \textbf{CB} & \textbf{SGI} & \textbf{FA} & \textbf{BA} & & & & \\
\hline

Chunyi \textit{et al.} \cite{Chunyi:2015} & Latency-aware rate adaptation mechanism & No & No & No & Yes & No & Frame aggregation helps to reduce end-to-end delay & IEEE 802.11n & 

 1) Reduces the ultimate latency for delay sensitive applications,
 2) Software/hardware retransmission dispatching
 & 
 1) Channel noise and interference are absent,
 2) Latency of different access categories is not discussed
\\
\hline

Hassan \textit{et al.} \cite{Hassan:2009} & Orthogonal multiple access scheme for different QoS requirements & Yes & No & No & No & No & MIMO-OFDMA improves QoS of the network & IEEE 802.11n & 

 1) Considers transmission bit-rates and powers for various channel conditions,
 2) Two different subcarrier sorting schemes,
 3) Downlink channel adaptation 
 & 
 1) Channel bonding is not considered,
 2) Does not discuss the placing of MIMO antenna
 \\
\hline

Ge \textit{et al.} \cite{Ge:2014} & Energy efficient frame aggregation scheme & No & No & No & Yes & No & Dynamic size of aggregated frame reduces packet transmission delay & IEEE 802.11n & 

 1) Change of aggregated frame size according to frame error rate, 
 2) Reduces energy consumption of channel contention
 & 
  1) Noise and interference are not considered, 
  2) Implementation details are not provided
 \\
\hline

Selvam \textit{et al.} \cite{Selvam:2010} & Frame aggregation scheduler & No & No & No & Yes & No & Dynamically chosen aggregated frame size reduces end-to-end delay & IEEE 802.11n & 

 1) Detailed simulation study of aggregation schemes, 
 2) Dynamically selects aggregated frame size and aggregation technique
 & 
  1) Different channel conditions are not considered, 
  2) Does not discuss implementation details
 \\
\hline

Anwar \textit{et al.} \cite{Anwar:2012} & Optimized A-MSDU aggregation with retransmission of subframe & No & No & No & Yes & No & Optimized A-MSDU frame header enhances VoIP performance & IEEE 802.11n & 

 1) An optimized header for A-MSDU frame, 
 2) Selective retransmission of subframes 
 & 
 1) Implementation details are not provided,
 2) Different channel conditions are not considered 
 \\
\hline

Chen \textit{et al.} \cite{Chen:2006} & Performance and implementation issues of MIMO-OFDM and OFDM-UWB & Yes & No & No & No & No & MIMO-OFDM improves network capacity & IEEE 802.11n & 

  Distinctive differences between OFDM-UWB and IEEE 802.11n technologies &

  Different channel signal strengths are not mentioned \\

\hline

Hajlaoui \textit{et al.} \cite{Jabri:2012} & Performance evaluation of IEEE 802.11n in real condition & No & No & No & Yes & No & Dynamic size of frame aggregation increases QoS and fairness of multimedia applications & IEEE 802.11n & 

 1) A frame aggregation scheduler,
 2) QoS requirements with multimedia applications,
 3) Dynamic adjustment of aggregated frame size,
 4) Access category of IEEE 802.11e
 & 
  Does not provide implementation details 
 \\
\hline

Lin \textit{et al.} \cite{Fang:2009} & Model to analyze capability of MIMO-OFDM of IEEE802.11n in Mobile Networking Architecture & Yes & No & No & No & No & TCP/IP capability is enhanced under multiple access point mode & IEEE 802.11n & 

 1) Vehicular ratio within the signal coverage,
 2) TCP/IP capability under multiple access points
 & 
 1) Noise and interference are not included,
 2) Does not discuss the placing of MIMO antennas  
 \\
\hline

Sriram \textit{et al.} \cite{Lakshmanan:2011} & New metric for selecting stream in MIMO & Yes & No & No & No & No & MIMO-based metric improves TCP/UDP throughput & IEEE 802.11n & 

 1) Intelligent rate selection mechanism can be developed,
 2) Experimental performance analysis of MIMO is carried out
 & 
 Implementaion and synchronization details of MIMO are not mentioned
 \\
\hline

\end{tabular} 
\label{table:5}
\end{table*}

\begin{table*}
\caption{Consideration of one enhanced feature of HT-WLANs (continue...)}
\centering
\begin{tabular}{|p{1cm}|p{1.6cm}|l|l|l|l|l|p{2cm}|p{1cm}|p{2.7cm}|p{2.7cm}|}
\hline
\textbf{Item} & \textbf{Objective} & \multicolumn{5}{c|}{\textbf{Trade-off}} & \textbf{Impact on upper layer protocols} & \textbf{Category} & \textbf{Strength} & \textbf{Weakness} \\
\cline{3-7} 
 & & \textbf{MIMO} & \textbf{CB} & \textbf{SGI} & \textbf{FA} & \textbf{BA} & & & & \\
\hline

Paleerat \textit{et al.} \cite{Paleerat:2014} & Performance analysis of different MIMO and beamforming techniques & Yes & No & No & No & No & MIMO and beamforming enhance TCP/UDP throughput & IEEE 802.11n & 

 Trade-off between both the schemes 
 & 
 1) Does not provide implementaion and synchronization details of MIMO,
 2) Effect of interference in their trade-offs is not discussed
 \\
\hline

Ting \textit{et al.} \cite{Ting:2011} & Coverage and capacity performance analysis of MIMO multihop network  & Yes & No & No & No & No & MIMO enriches end-to-end network capacity & IEEE 802.11n & 

 1) Performance analysis in typical rural area, 
 2) Different design parameters of multiradio
 & 
 1) Throughput maximization is absent,
 2) Different mobility conditions are not considered
 \\
\hline

Asuman \textit{et al.} \cite{Yavanoglu:2011} & Performance analysis of spatial multiplexing of MIMO-OFDM & Yes & No & No & No & No & MIMO-OFDM enhances network capacity & IEEE 802.11n & 

 1) Consideration of data rate and compactness, 
 2) Linear low-complexity MIMO detectors
 & 
 1) Synchronization of MIMO spatial streams is not discussed,
 2) Trade-offs between different MIMO techniques are absent
 \\
\hline

Lin \textit{et al.} \cite{Chang:2010} & Channel estimation algorithm & Yes & No & No & No & No & MIMO with channel estimation reduces packet transmission latency & IEEE 802.11n & 

 1) Increases performance of MIMO channel estimation, 
 2) Decreasing of decoding rate at receiver side,
 3) Experimentally implemented and verified
 & 
 1) Different channel conditions are not considered,
 2) Designing aspects are absent
 \\
\hline

Lu \textit{et al.} \cite{Tai:2007} & Impact of improved channel estimation & Yes & No & No & No & No & MIMO with improved channel estimation reduces end-to-end delay & IEEE 802.11n & 

 1) Consideration of three channel estimation methods, 
 2) Consideration of transmit beamforming
 & 
  Detailed impact analysis with trade-offs is not discussed
 \\

\hline

Daniel \textit{et al.} \cite{Daniel:2012} & Frame aggregation scheduler & No & No & No & Yes & No & Improves QoS and power consumption of overall network & IEEE 802.11n & 

 1) Power saving mode, 
 2) Consideration of Congestion in the network 
 & 
 1) Considers only MAC layer information, 
 2) Does not consider PHY enhanced features
 \\
\hline

Muck \textit{et al.} \cite{Muck:2006} & Optimizing the mapping of irregular and systematic low density parity check (LDPC) code onto OFDM & Yes & No & No & No & No & Improves overall network capacity & IEEE 802.11n & 

 1) Both SISO and MIMO are considered, 
 2) Adaptive interleaver optimized for given channel impulse
 & 
 1) Trade-offs between SISO and MIMO are not provided,
 2) Synchronization and implementation aspects are not given
 \\
\hline

Ali \textit{et al.} \cite{Mahmood:2013} & Performance analysis of MIMO OFDM time synchronization & Yes & No & No & No & No & Synchronized MIMO reduces end-to-end delay & IEEE 802.11n & 

 Synchronization of MIMO spatial streams 
 & 
 1) MIMO antenna placement is not considered,
 2) Noise and interference levels are not considered
 \\
\hline

Ertug \textit{et al.} \cite{Ertug:2011} & Analysis of MIMO antenna array & Yes & No & No & No & No & MIMO reduces packet transmission delay & IEEE 802.11n & 

 1) Higher order space multimode polarization diversity, 
 2) Use of higher number of antenna ports in lower space,
 3) Low cross-polar discrimination values between transmit and receive antennas
 & 
 1) Synchronization between different antennas is not mentioned,
 2) Trade-offs between different parameters are absent
 \\
\hline

Min \textit{et al.} \cite{Min:2008} & Iterative SNR estimator for using in link adaptation & Yes & No & No & No & No & Achieves end-to-end throughput gain & IEEE 802.11n & 

 1) Extends to MIMO model,
 2) Dynamic channel condition for link adaptation
 & 
 1) Does not provide implementation details of the estimator,
 2) Interference is not considered
 \\
\hline

\end{tabular} 
\label{table:6}
\end{table*}

\begin{table*}
\caption{Consideration of one enhanced feature of HT-WLANs (continue...)}
\centering
\begin{tabular}{|p{1cm}|p{1.6cm}|l|l|l|l|l|p{2cm}|p{1cm}|p{2.7cm}|p{2.7cm}|}
\hline
\textbf{Item} & \textbf{Objective} & \multicolumn{5}{c|}{\textbf{Trade-off}} & \textbf{Impact on upper layer protocols} & \textbf{Category} & \textbf{Strength} & \textbf{Weakness} \\
\cline{3-7} 
 & & \textbf{MIMO} & \textbf{CB} & \textbf{SGI} & \textbf{FA} & \textbf{BA} & & & & \\
\hline

Xi \textit{et al.} \cite{Xi:2008} & MIMO-based link adaptation & Yes & No & No & No & No & MIMO with channel state information improves TCP/UDP throughput & IEEE 802.11n & 

 1) Cross-layer design,
 2) Employs channel state information
 & 
 1) Link adaptation with only MIMO of PHY/MAC enhancement,
 2) Does not mention synchronization between antennas with different channel conditions
 \\
\hline

Marcio \textit{et al.} \cite{Marcio:2013} & Performance analysis of frame aggregation with video streaming & No & No & No & Yes & No & Increases QoS of video application & IEEE 802.11n & 

 1) Trade-offs between A-MSDU and A-MPDU,
 2) Use of video streaming,
 3) Suggestion of QoS parameters
 & 
 1) Detailed analysis of QoS is not present,
 2) Video streaming without mobility
 \\
\hline

Yang \textit{et al.} \cite{Yang:2004} & Proposal of MAC strategies to overcome fundamental overhead and improve performance & No & No & No & Yes & No & Reduces end-to-end packet transmission latency & IEEE 802.11n & 

 1) Packing and concatenation of frames,
 2) Multiple frame transmission
 & 
 1) Lack of BA mechanism to overcome MAC overhead,
 2) Implementation details with mobility is not discussed
 \\
\hline

Tassiulas \textit{et al.} \cite{Tassiulas:2008} & High throughput multicast transmission & Yes & No & No & No & No & Provides low latency for real time applications & IEEE 802.11n & 

 1) Near real time multimedia content transmission,
 2) Enables beamforming
 & 
  channel error condition is not considered
 \\
\hline

Xia \textit{et al.} \cite{Xia:2009} & Open-loop link adaptation mechanism & Yes & No & No & No & No & Reduces user-level response delay and protocol overhead & IEEE 802.11n & 

 1) Modified link estimation and probing behavior, 
 2) Detailed design guidelines 
 & 
 1) Does not consider other enhanced PHY features,
 2) Involvement of MAC layer is absent
 \\
\hline

Chi-Yu \textit{et al.} \cite{Chi-Yu:2015} & Energy efficiency analysis of rate adaptation & Yes & No & No & No & No & MIMO enhances goodput & IEEE 802.11n & 

 1) Energy efficiency of MIMO rate adaptation,
 2) Trade-offs between goodput and energy
 & 
 1) Energy efficiency with different channel conditions is not mentioned,
 2) Energy efficieny with mobility is not discussed
 \\
\hline

Richard \textit{et al.} \cite{Richard:2006} & Performance analysis of MIMO-OFDM standard & Yes & No & No & No & No & Increases range for a transmission delay and thus enhances end-to-end throughput & IEEE 802.11n & 

 1) Throughput enhancement using MIMO-OFDM,
 2) Throughput comparison with legacy standards
 & 
 1) Limitation of MIMO-OFDM is not mentioned,
 2) Different network conditions are not focused
 \\
\hline

Mohamed \textit{et al.} \cite{Mohamed:2014} & Performance evaluation of different rate adaptation algorithms & No & No & No & Yes & No & Different PHY/MAC parameters help to improve TCP/UDP throughput & IEEE 802.11n & 

 1) Comparative study of different algorithms,
 2) Simulation of different algorithms
 & 
 Detailed description of each algorithm is not presented
 \\
\hline

Michael \textit{et al.} \cite{Michael:2014} & Providing high bandwidth back-hauling with minimal power consumption & Yes & No & No & No & No & Provides low network latency and high network capacity & IEEE 802.11n & 

 1) Low-power cost-effective MIMO hardware,
 2) Single antenna with cross polarization 
 & 
 1) Does not provide the design issue of MIMO antenna,
 2) Implementation details are not mentioned,
 3) Coexistence of channel bonding is not included
 \\
\hline

Feng \textit{et al.} \cite{Feng:2010} & Link adaptation mechanism based on frame aggregation & No & No & No & Yes & No & Dynamic setting of parameters improves TCP/UDP throughput & IEEE 802.11n & 

 1) Dynamic channel condition,
 2) Dynamic adjustment of system parameters, 
 3) Consideration of MCS values and noise levels
 & 
 1) Mobile environment is not included,
 2) Analysis with several network parameters is absent
 \\
\hline
\end{tabular} 
\label{table:7}
\end{table*}

\begin{table*}[t]
\caption{Consideration of one enhanced feature of HT-WLANs (continue...)}
\centering
\begin{tabular}{|p{1cm}|p{1.6cm}|l|l|l|l|l|p{2cm}|p{1cm}|p{2.7cm}|p{2.7cm}|}
\hline
\textbf{Item} & \textbf{Objective} & \multicolumn{5}{c|}{\textbf{Trade-off}} & \textbf{Impact on upper layer protocols} & \textbf{Category} & \textbf{Strength} & \textbf{Weakness} \\
\cline{3-7} 
 & & \textbf{MIMO} & \textbf{CB} & \textbf{SGI} & \textbf{FA} & \textbf{BA} & & & & \\
\hline

Yun \textit{et al.} \cite{Yun:2010} & MIMO based rate adaptation & Yes & No & No & No & No & MIMO helps to increase application level throughput & IEEE 802.11n & 

 1) Rate options between intra-mode and inter-mode, 
 2) Consideration of collision settings and mobility   
 & 
 1) Interaction details between MAC and PHY layer are not mentioned,
 2) Lack of performance analysis of different upper layer parameters 
 \\
\hline

Federico \textit{et al.} \cite{Federico:2015} & Analyzation and effective parameter set tuning for rate adaptation algorithm & Yes & No & No & No & No & Enhances application level network capacity & IEEE 802.11n & 

 Communication system deployed in industrial scenario 
 & 
 1) Design of MIMO stream is not focused,
 2) Considers only MIMO as PHY enhancement 
 \\
\hline

Jilong \textit{et al.} \cite{Jilong:2015} & Optimized two-level frame aggregation & No & No & No & Yes & No & Dynamic adjustment of frame aggregation reduces end-to-end packet transmission delay  & IEEE 802.11n & 

 1) Optimized aggregation level,  
 2) Dynamically adjusted aggregation level according to subframe size
 & 
 1) Implementation details are absent,
 2) Different channel conditions are not focused 
 \\
\hline
\end{tabular} 
\label{table:71}
\end{table*}

\begin{table*}
\caption{Consideration of two enhanced features of HT-WLANs}
\centering
\begin{tabular}{|p{1cm}|p{1.6cm}|l|l|l|l|l|p{2cm}|p{1cm}|p{2.7cm}|p{2.7cm}|}
\hline
\textbf{Item} & \textbf{Objective} & \multicolumn{5}{c|}{\textbf{Trade-off}} & \textbf{Impact on upper layer protocols} & \textbf{Category} & \textbf{Strength} & \textbf{Weakness} \\
\cline{3-7} 
 & & \textbf{MIMO} & \textbf{CB} & \textbf{SGI} & \textbf{FA} & \textbf{BA} & & & & \\
\hline

Anwar \textit{et al.} \cite{Anwar:2014} & A framework for packet loss differentiation & No & No & No & Yes & Yes & Use of loss pattern reduces application level transmission delay & IEEE 802.11n & 

 1) Considers noise, collisions, and hidden nodes,
 2) Implementation in ath9k driver
 & 
 1) Performance is not analyzed in IEEE 802.11ac,
 2) Does not consider PHY enhancements
 \\
\hline

Frohn \textit{et al.} \cite{Frohn:2011} & Characterization of effective throughput in multi-hop environment & No & No & No & Yes & Yes & Increases application level throughput & IEEE 802.11n & 

 1) Analytical model to capture the effects of FA and BACK,
 2) Considers bit error rate, aggregation level and path length,
 3) Comparative study of mathematical and simulation results
 & 
 Noise and interference level are not mentioned
 \\
\hline

Kolap \textit{et al.} \cite{Kolap:2012} & Frame aggregation mechanism & No & No & No & Yes & Yes & Appropriate size of frame aggregation reduces end-to-end latency & IEEE 802.11n & 
  Analyzes two-level frame aggregation &
  Channel error condition is not considered \\
\hline

Bakeel \textit{et al.} \cite{Baba:2012} & Real time frame aggregation scheduler & No & No & No & Yes & Yes & Enhances QoS of VoIP, videos and interactive gaming & IEEE 802.11n & 
 Enables retransmission in real-time A-MSDU aggregation scheduler &
  Channel error condition is not discussed \\
\hline

Wang \textit{et al.} \cite{Wang:2009} & Examining network performance enhancement by IEEE 802.11n MAC layer features & No & No & No & Yes & Yes & MAC enhancements improve VoIP performance & IEEE 802.11n & 

 1) Implements a new IEEE 802.11n module in the NS-2, 
 2) VoIP performance evaluation,
 3) Considers reverse direction mechanism
 & 
 Different channel conditions are not considered 
 \\
\hline

Nakajima \textit{et al.} \cite{Nakajima:2005} & New BACK mechanism  & No & No & No & Yes & Yes & Reduces end-to-end transmission delay & IEEE 802.11n & 
 Compressed BACK as a selective repeat mechanism &
  Can not be applicable when fragmentation is used in frame aggregation\\
\hline

Deek \textit{et al.} \cite{Deek:2014} & Efficient channel bonding technique to fully utilize $40$ MHz channel & Yes & Yes & No & No & No & Appropriate channel bonding increases TCP/UDP throughput & IEEE 802.11n & 

 1) Identifies the factors that influence performance of channel bonding, 
 2) Intelligent channel bonding, 
 3) Incorporation of MIMO
 & 
  Does not provide implementation details of a network detector to identify interference
 \\
\hline

Alsahlany \textit{et al.} \cite{Alsahlany:2011} & MAC performance in multipacket reception with multiple antenna system for fiber network & Yes & No & No & Yes & No & MIMO reduces application level propagation delay & IEEE 802.11n & 

 1) Derivation of closed-form expression to maximize throughput, 
 2) Considers network size and fiber propagation delay,
 3) Comparative study of ideal and error prone channel
 & 
 1) Detailed analysis of selected parameter set is not presented,
 2) Mobility is not considered
\\
\hline

Deek \textit{et al.} \cite{Deek:2013} & Rate adaptation technique & Yes & Yes & No & No & No & MIMO with channel bonding increases network capacity & IEEE 802.11n & 
 1) Closed-loop rate adaptation, 
 2) Adapts with channel bonding, 
 3) MIMO transmission technique,
 4) Implements on Atheros-based devices
 & 
  Time bound analysis is not mentioned
 \\
\hline

Kriara \textit{et al.} \cite{sample:2015} & Sender-side link adaptation algorithm based on RSSI & Yes & Yes & No & No & No & MIMO and channel bonding do efficient rate selection and increase TCP/UDP throughput & IEEE 802.11n & 

 1) Sender selects data rate without receiver's acknowledgement, 
 2) Different advanced MCS values 
 & 
 1) Threshold-based mechanism,
 2) Noise and interference are not considered
 \\
\hline

Letor \textit{et al.} \cite{Letor:2012} & Feasibility of employing IEEE 802.11n to multimedia multicast real-time content & Yes & Yes & No & No & No & Improves TCP/UDP throughput of real-time application & IEEE 802.11n & 

 Mutimedia real-time transmission 
 & 
 1) Does not consider traffic categories, 
 2) Different channel conditions are not provided
\\
\hline

\end{tabular} 
\label{table:8}
\end{table*}

\begin{table*}
\caption{Consideration of two enhanced features of HT-WLANs (Continue...)}
\centering
\begin{tabular}{|p{1cm}|p{1.6cm}|l|l|l|l|l|p{2cm}|p{1cm}|p{2.7cm}|p{2.7cm}|}
\hline
\textbf{Item} & \textbf{Objective} & \multicolumn{5}{c|}{\textbf{Trade-off}} & \textbf{Impact on upper layer protocols} & \textbf{Category} & \textbf{Strength} & \textbf{Weakness} \\
\cline{3-7} 
 & & \textbf{MIMO} & \textbf{CB} & \textbf{SGI} & \textbf{FA} & \textbf{BA} & & & & \\
\hline

Nakajima \textit{et al.} \cite{Nakajima:2007} & Enhancement of the performance of BACK & No & No & No & Yes & Yes & Helps to reduce packet transmission latency & IEEE 802.11n & 
 Extended BACK mechanism to be adapted to frame aggregation &
  Channel error condition is not considered \\
\hline

Yoriko \textit{et al.} \cite{Yoriko:2007} & Mechanism to control coexistence of $20/40$ MHz bandwidth & Yes & Yes & No & No & No & Enhances TCP/UDP network capacity & IEEE 802.11n & 

 1) MAC proposal,
 2) Dynamically controls $20/40$ MHz bandwidth levels 
 & 
 1) Implementation details are not given, 
 2) Noise and interference levels are not mentioned
 \\
\hline

Abu \textit{et al.} \cite{Abu:2011} & Multi-rate performance anomaly with frame aggregation and BACK & No & No & No & Yes & Yes & Multi-rate performance anomaly increases packet transmission delay & IEEE 802.11n & 
  Impact of multi-rate performance anomaly with FA and BACK &
  Explanation of trade-offs between different MAC features is absent\\
\hline

Kwong \textit{et al.} \cite{Kwong:2012} & Design and performance analysis of multihop multiradio network & Yes & Yes & No & No & No & Achieves gain in TCP/UDP throughput & IEEE 802.11n & 

 1) Choice of MCS level,
 2) Backhaul link rate,
 3) Multihop chain distance
 & 
 1) Does not discuss trade-offs between channel bonding and spatial streams,
 2) Different channel conditions are absent
 \\
\hline

Yao \textit{et al.} \cite{Xu:2008} & Highly integrated MIMO & Yes & Yes & No & No & No & Reduces application level propagation delay & IEEE 802.11n & 

 1) Design of access point radio, 
 2) Small size and low cost
 & 
 1) Does not provide implementation and synchronization details of MIMO, 
 2) Other PHY enhancements are absent
 \\
\hline

Pelechrinis \textit{et al.} \cite{Pel:2010} & Link quality investigation at high data rates & Yes & Yes & No & No & No & MIMO with channel bonding increases end-to end performance under different channel conditions & IEEE 802.11n & 

 1) Impact of packet size,
 2) Different channel conditions 
 & 
 1) Effect of each PHY/MAC enhanced feature is absent, 
 2) Mobility is not mentioned
 \\
\hline

Mustafa \textit{et al.} \cite{Mustafa:2012} & Interaction between TCP and MAC frame aggregation & No & No & No & Yes & Yes & MAC enhancements improve the performance of TCP congestion control algorithms and increase packet transmission rate & IEEE 802.11n & 

 1) End-to-end transfer rate,
 2) BACK with frame aggregation, 
 3) Different TCP congestion algorithms,
 4) Access category of IEEE 802.11e
 & 
 Different channel conditions are not included 
 \\
\hline



Yao \textit{et al.} \cite{Yao:2012} & Channel Sensing mechanism for IEEE 802.11n MIMO device & Yes & Yes & No & No & No & Helps to enhance network capacity & IEEE 802.11n & 

 Energy detection and carrier sensing to detect channel state
& 
 1) Implementation details are not discussed,
 2) Discussion restricts upto $40$ MHz
\\
\hline

Santos \textit{et al.} \cite{Santos:2011} & Performance evaluation of high throughput MAC mechanisms & No & No & No & Yes & Yes & MAC enhancements improve TCP/UDP throughput & IEEE 802.11n & 

 1) Examines frame aggregation, BA and reverse direction,
 2) Network performance of each parameter
 & 
 1) Performance analysis with respect to only number of stations,
 2) Only throughput is measured
\\
\hline

Sandip \textit{et al.} \cite{Chakraborty:2015} & Mitigates hidden node and exposed node problem & No & No & No & Yes & Yes & Enhances upper layer performance & IEEE 802.11n and IEEE 802.11ac & 

 1) Exploits frame aggregation and BACK,
 2) Opportunistic collision avoidance method 
 & 
 1) Trade-offs between frame aggregation and BACK are not mentioned,
 2) Implementation details are absent
\ \\
\hline

Pravati \textit{et al.} \cite{Pravati:2016} & Modeling the impact of channel errors over the MAC channel access  & No & No & No & Yes & Yes & Improves QoS of several access categories & IEEE 802.11n & 

 1) Exploits both frame aggregation and BACK,
 2) QoS with different access categories,
 3) Use of Markov model
 & 
 1) Insufficient explanation of BACK in the model,
 2) Evaluation under mobile environment is absent
 \\
\hline
\end{tabular} 
\label{table:9}
\end{table*}


\begin{table*}
\caption{Consideration of three enhanced features of HT-WLANs}
\centering
\begin{tabular}{|p{1cm}|p{1.6cm}|l|l|l|l|l|p{2cm}|p{1cm}|p{2.7cm}|p{2.7cm}|}
\hline
\textbf{Item} & \textbf{Objective} & \multicolumn{5}{c|}{\textbf{Trade-off}} & \textbf{Impact on upper layer protocols} & \textbf{Category} & \textbf{Strength} & \textbf{Weakness} \\
\cline{3-7} 
 & & \textbf{MIMO} & \textbf{CB} & \textbf{SGI} & \textbf{FA} & \textbf{BA} & & & & \\
\hline

Janis \textit{et al.} \cite{Janis:2012} & Basic performance analysis of IEEE 802.11g/n & Yes & Yes & No & Yes & No & Enhances TCP/UDP network capacity & IEEE 802.11n & 

 1) Compares energy consumption,
 2) Effect of mobility,
 3) SNR is considered
 & 
 Trade-off analysis of several features of IEEE 802.11n is not provided
\\
\hline

Ghaleb \textit{et al.} \cite{Ghaleb:2012} & Performance analysis in terms of throughput of IEEE 802.11n & Yes & Yes & No & No & Yes & Higher MCS values help to increase TCP/UDP throughput & IEEE 802.11n & 

 1) Use of OPNET Modeler,
 2) Consideration of MCS values, 
 3) Different access categories are considered,
 4) Effect of size of IP packet 
 & 
 1) All PHY/MAC enhanced features are not included,
 2) Various channel conditions are absent
 \\
\hline

Sean \textit{et al.} \cite{Biswas:2007} & Analysis of VoIP for high throughput transmission & Yes & Yes & No & Yes & No & Enhances QoS of end-to-end transmission & IEEE 802.11n & 

 1) Consideration of QoS packet transmission,
 2) Downlink and uplink transmission,
 3) Power save multi-poll mechanism
 & 
 1) All access categories are not mentioned,
 2) Absence of trade-offs between several included parameters
 \\
\hline

Sandip \textit{et al.} \cite{Sandip:2014} & TCP performance evaluation in burst and correlated losses in high data rate communication & No & Yes & No & Yes & Yes & Achieves gain in TCP throughput for different congestion control schemes & IEEE 802.11n & 

1) Four different TCP variants,
2) Wireless mesh backbone,
3) Considers channel bonding with MAC enhancements
 & 
 Different interference and noise levels are absent
 \\
\hline

Karmakar \textit{et al.} \cite{karmakar2015dynamic} & Design of a dynamic link adaptation mechanism for HT-WLANs & No & Yes & Yes & Yes & No & Increases TCP throughput and decreases packet loss and delay compared to the other competing schemes & IEEE 802.11n and IEEE 802.11ac & 

1) Use of statistic table to store the past information,
2) Creation of configuration set,
3) Consideration of MCS values,
4) State transition model followed by the algorithm
 & 
1) Does not consider noise power of the channel
2) Large size of statistic table for long run.
 \\ 
\hline
\end{tabular} 
\label{table:10}
\end{table*}


\begin{table*}
\caption{Consideration of all enhanced features of HT-WLANs}
\centering
\begin{tabular}{|p{1cm}|p{1.6cm}|l|l|l|l|l|p{2cm}|p{1cm}|p{2.7cm}|p{2.7cm}|}
\hline
\textbf{Item} & \textbf{Objective} & \multicolumn{5}{c|}{\textbf{Trade-off}} & \textbf{Impact on upper layer protocols} & \textbf{Category} & \textbf{Strength} & \textbf{Weakness} \\
\cline{3-7} 
 & & \textbf{MIMO} & \textbf{CB} & \textbf{SGI} & \textbf{FA} & \textbf{BA} & & & & \\
\hline

Singh \textit{et al.} \cite{Singh:2015} & Performance optimization of QoS based DCF-IEEE 802.11n WLAN & Yes & Yes & Yes & Yes & Yes & Enhances QoS of network performance & IEEE 802.11n & 

 1) Considers QoS, 
 2) Random RTS threshold
 & 
  1) New QoS feature is not considered,
  2) Enhanced PHY/MAC parameter setting is absent
 \\
\hline

Sotenga \textit{et al.} \cite{Sotenga:2015} & Theoretical throughput estimation for several access categories under non saturated conditions & Yes & Yes & Yes & Yes & Yes & All PHY/MAC enhancements improve QoS and fairness of upper layer performance & IEEE 802.11n & 

 1) Fairness linked with QoS,  
 2) Compares analytical result with practical result
 & 
 QoS with different enhanced PHY/MAC features is not mentioned
 \\
\hline

Sandip \textit{et al.} \cite{Sandip:2016} & Estimation and sampling scheme to optimally configure PHY/MAC enhancements & Yes & Yes & Yes & Yes & Yes & Dynamic parameter setting increases TCP/UDP throughput significantly & IEEE 802.11n and IEEE 802.11ac & 

 1) Consideration of the maximum PHY/MAC enhancements,  
 2) Dynamic tuning of parameter set
 & 
 1) Time complexity of proposed scheme is not explained,
 2) Detailed application of the scheme is not mentioned
 \\
\hline

Letor \textit{et al.} \cite{Letor:2012} & Performance analysis of multimedia multicast transmission & Yes & Yes & Yes & Yes & Yes & Increases application level performance of multimedia transmission & IEEE 802.11n & 

 1) Experimentally analyzes the performance of available IEEE 802.11n hardware, 
 2) Comparison with standard multicast scheme 
 & 
  1) Does not specify the detailed settings of several PHY/MAC enhanced features,
  2) Analysis of different network parameters is not included
 \\
\hline

Charfi \textit{et al.} \cite{Charfi:2013} & A survey of QoS and PHY/MAC enhancements of high throughput WLANs & Yes & Yes & Yes & Yes & Yes & Different PHY/MAC enhancements increase QoS of multimedia applications & Both IEEE 802.11n and IEEE 802.11ac & 

 1) Study of different PHY/MAC enhancements, 
 2) Major MAC mechanisms for improving network performance,
 3) QoS for real time multimedia transmission 
 & 
  1) Detailed analysis of PHY enhancements for improving network performance is absent,
  2) Does not provide trade-off analysis of several PHY/MAC enhanced features,
  3) Performance examination of different network parameters is absent
 \\
\hline

Nobuo \textit{et al.} \cite{Nobuo:2013} & Modified active AP selection scheme with link speed change in wireless mesh network & Yes & Yes & Yes & Yes & Yes & Reduces transmission delay and enhances upper layer performance  & IEEE 802.11n & 

 1) New design parameter set for wireless mesh network, 
 2) Selects link with sufficient throughput
 & 
  1) Selection of different enhanced PHY/MAC features are not mentioned,
  2) Dynamic channel condition is not considered,
  3) Mobile environment is not focused
 \\
\hline

Shimajiri \textit{et al.} \cite{Hayashi:2009} & Discussion of RTL design of high-throughput Low Density Parity Check (LDPC) decoder in IEEE 802.11n & Yes & Yes & Yes & Yes & Yes & Enhances reliability of multimedia access & IEEE 802.11n & 

 1) Performance achievement close to Shannon limit, 
 2) RTL design using min-sum algorithm
 & 
  1) Settings of several PHY/MAC enhanced parameter set are not focused,
  2) Different network scenarios are not considered,
  3) Dynamic channel condition is not properly explained
 \\
\hline
\end{tabular} 
\label{table:11}
\end{table*}

\begin{table*}
\caption{Consideration of all enhanced features of HT-WLANs (continue...)}
\centering
\begin{tabular}{|p{1cm}|p{1.6cm}|l|l|l|l|l|p{2cm}|p{1cm}|p{2.7cm}|p{2.7cm}|}
\hline
\textbf{Item} & \textbf{Objective} & \multicolumn{5}{c|}{\textbf{Trade-off}} & \textbf{Impact on upper layer protocols} & \textbf{Category} & \textbf{Strength} & \textbf{Weakness} \\
\cline{3-7} 
 & & \textbf{MIMO} & \textbf{CB} & \textbf{SGI} & \textbf{FA} & \textbf{BA} & & & & \\
\hline

Toru \textit{et al.} \cite{Toru:2013} & Modified algorithm for routing tree considering large speed change to adopt IEEE 802.11n & Yes & Yes & Yes & Yes & Yes & Reduces routing performance latency & IEEE 802.11n & 

 1) Selection of link between host and the corresponding AP, 
 2) Distance speed relationship considering signal quality,
 3) Different topologies for performance evaluation
 & 
  1) Noise and interference levels are not mentioned,
  2) Settings of several PHY/MAC enhanced parameter set are not discussed,
  3) Dynamic channel condition is not properly explained
 \\
\hline

Tajima \textit{et al.} \cite{Tajima:2013} & Extended clustering algorithm with consideration of link speed in mesh networks & Yes & Yes & Yes & Yes & Yes & Considering link speed, all PHY/MAC enhancements improve upper layer performance & IEEE 802.11n & 

 1) Trade-offs between link distance and interference,
 2) Simulations in grid topology 
& 
  1) Detailed PHY/MAC enhanced parameter settings are not given,
  2) Does not provide performance for several network parameters,
  3) Lack of different channel conditions 
 \\
\hline

Funabiki  \textit{et al.} \cite{Funa:2012} & Proposal of a routing algorithm for wireless mobile network considering the changes of link speed according to distance & Yes & Yes & Yes & Yes & Yes & Reduces routing delay and helps to increase end-to-end throughput & IEEE 802.11n & 

 1) Trade-offs between throughput and link distance,
 2) Simulations with different topologies 
 & 
  1) Detailed PHY/MAC enhanced parameter settings are not mentioned,
  2) Does not mention performance for several network parameters,
  3) Dynamic channel condition is not explained
 \\
\hline

Zawar \textit{et al.} \cite{Zawar:2015} & Comparative performance evaluation of IEEE 802.11n and IEEE 802.11ac & Yes & Yes & Yes & Yes & Yes & Both can achieve gain in TCP/UDP throughput & IEEE 802.11n and IEEE 802.11ac & 

 1) Real indoor environment, 
 2) Line of sight and non-line of sight conditions, 
 3) Presence of interference 
 & 
  1) Configuration of enhanced PHY/MAC parameters is not mentioned,
  2) Advantage of new IEEE 802.11ac features is not explained
 \\
\hline

\end{tabular} 
\label{table:12}
\end{table*}


\begin{table*}
\caption{Comparative study of IEEE 802.11ac related works}
\centering
\begin{tabular}{|p{1cm}|p{1.7cm}|l|l|p{1cm}|l|l|l|p{1.8cm}|p{2.6cm}|p{2.5cm}|}
\hline
\textbf{Item} & \textbf{Objective} & \multicolumn{6}{c|}{\textbf{Trade-off}} & \textbf{Impact on upper layer protocols} & \textbf{Strength} & \textbf{Weakness} \\
\cline{3-8} 
 & & \textbf{MU-MIMO} & \textbf{CB} & \textbf{Dynamic bandwidth management} & \textbf{SGI} & \textbf{FA} & \textbf{BA} & & & \\
\hline

Eldad  \textit{et al.} \cite{Eldad:2011} & An overview of IEEE Gigabit WLANs amendments (IEEE 802.11ac) & Yes & Yes & Yes & No & Yes & No & Improves network capacity & 

 1) Detailed description of enhanced PHY/MAC features,
 2) Consideration of beamforming scheme
 & 
 1) Does not provide performance trade-offs between several PHY/MAC enhanced features,
 2) Uper layer performance study is not mentioned
 \\
\hline

Juliet  \textit{et al.} \cite{Femi:2013} & Empirical measurements of an IEEE 802.11ac system based on the beamforming technology & Yes & No & No & No & No & No & Beamforming reduces end-to-end transmission delay & 

 1) Determines effectiveness of beamforming scheme,
 2) Investigation in indoor wireless links
 & 
 Other enhanced PHY/MAC features are not included 
\\
\hline

Dianu \textit{et al.} \cite{Dia:2014} & Measurement-based analysis of IEEE 802.11ac in indoor environment & Yes & Yes & No & No & No & No & MU-MIMO with channel bonding improves application level performance & 

 1) Measurement in office building,
 2) Different channel conditions
 & 
 1) Performance trade-offs between different PHY/MAC enhancements are not given,
 2) Detailed analysis of performance enhancements is absent
 \\
\hline

Boris \textit{et al.} \cite{Boris:2012} & Analyzation of packet aggregation method in MU-MIMO IEEE 802.11ac WLANs & Yes & No & No & No & Yes & No & Reduces end-to-end transmission latency  & 

 1) Effect of MU-MIMO over frame aggregation,
 2) Consideration of number of active stations,
 3) Impact of queue size
 & 
 1) Lack of different channel conditions with interference levels,
 2) Upper layer performance analysis is absent
 \\
\hline

Liao \textit{et al.} \cite{Liao:2013} & Evaluation of the features of IEEE 802.11ac in a wireless mesh network & Yes & Yes & No & No & Yes & No & Different PHY/MAC features help to increase upper layer performance & 

 1) Survey of different IEEE 802.11ac features,
 2) Evaluation of explicit compressed feedback (ECFB) scheme,
 3) Extended RTS/CTS that integrates ECFB 
 & 
 1) Coexistence of different PHY/MAC features is not explained, 
 2) Different network scenarios are absent
 \\
\hline

Bellalta \textit{et al.} \cite{Faridi:2014} & Performance analysis of channel bonding for short-range WLANs & No & Yes & No & No & No & No & Channel bonding enhances TCP/UDP throughput & 

 1) External interference,
 2) Analytical model and evaluation of performance
 & 
 1) Does not specify trade-offs between two considered channel access schemes,
 2) Detailed explanation of performance is not provided
 \\
\hline

Park \textit{et al.} \cite{Park:2011} & Performance analysis of dynamic channel switching between available bandwidths & No & Yes & Yes & No & No & No & Dynamic channel access improves TCP/UDP performance & 

 1) Study of dynamic and static channel access schemes,
 2) Consideration of clear channel assessment
 & 
 1) Noise and interference levels are not focused,
 2) Trade-offs between different bonding levels are not mentioned
 \\
\hline

Per \textit{et al.} \cite{Per:2014} & Implementation of interference alignment with joint transmission coordinated multipoint & Yes & No & No & No & No & No & Considering different MCS levels, MU-MIMO can further increase TCP/UDP performance & 

 1) Feedback compression,
 2) Different MCS levels
 & 
 Lack of the consideration of all PHY/MAC enhancements
 \\
\hline

\end{tabular} 
\label{table:13}
\end{table*}

\begin{table*}
\caption{Comparative study of IEEE 802.11ac related works (continue...)}
\centering
\begin{tabular}{|p{1.1cm}|p{1.7cm}|l|l|p{1cm}|l|l|l|p{1.8cm}|p{2.5cm}|p{2.5cm}|}
\hline
\textbf{Item} & \textbf{Objective} & \multicolumn{6}{c|}{\textbf{Trade-off}} & \textbf{Impact on upper layer protocols} & \textbf{Strength} & \textbf{Weakness} \\
\cline{3-8} 
 & & \textbf{MU-MIMO} & \textbf{CB} & \textbf{Dynamic bandwidth management} & \textbf{SGI} & \textbf{FA} & \textbf{BA} & & & \\
\hline

Zhu \textit{et al.} \cite{Zhu:2011} & Reports the current status of MU-MIMO considering the MAC enhancements needed for downlink MU-MIMO & Yes & No & No & No & No & No & Reduces packet transmission delay  & 

 1) Focusing on MAC modifications for downlink MU-MIMO,
 2) TXOP sharing with related issues
 & 
 Internal trade-offs among the modifications are not explained
 \\
\hline

Ong \textit{et al.} \cite{Ong:2011} & Mandatory and optional PHY and MAC enhanced features of IEEE 802.11ac & Yes & Yes & No & No & Yes & No & Enhances network capacity & 

 1) A hybrid frame aggregation method,
 2) Coexistence of defferent channel bonding levels
 & 
 1) Bonding levels without dynamic channel management,
 2) MU-MIMO is not focused
 \\
\hline

Hart \textit{et al.} \cite{Hart:2011} & Identification of problems regarding channel bonding and designs a MAC protection technique & No & Yes & Yes & No & No & No & Enhances fairness in application level & 

 1) MAC protection to overcome hidden node problem,
 2) Evaluates fairness
 & 
 1) Dynamic bandwidth management with higher bonding is not considered, 
 2) Lack of different channel conditions
 \\
\hline

Swetank \textit{et al.} \cite{Swetank:2015} & Experimental study of IEEE 802.11n/ac power consumption and throughput in modern smartphones & Yes & Yes & No & No & Yes & No & Increases TCP/UDP throughput and enhances network fairness & 

  1) Experiment with different types of smartphones,
  2) Trade-offs between throughput and power consumption,
  3) Guides to design power optimization algorithm,
  4) Evaluates fairness
 & 
  1) Does not focus on trade-offs among different PHY/MAC enhanced features,
  2) All PHY/MAC enhancements are not included
 \\
\hline

Zeng \textit{et al.} \cite{Prasant:2014} & Impact of channel width and number of spatial streams over energy efficiency and interference & Yes & Yes & No & No & No & No & Higher MCS values increase TCP/UDP throughput & 

  1) Different MCS levels,
  2) Trade-offs between considered features and power consumption,
  3) Benefits and challenges of using 802.11ac in practice,
  4) Unfairness issues
 & 
 1) All PHY/MAC enhancements are not included,
 2) Does not discuss trade-offs among different PHY/MAC enhanced features and power consumption
 \\
\hline

Cheng \textit{et al.} \cite{Rung-Shiang:2015} & Evaluation of performance of stream control transmission protocol (SCTP) over IEEE 802.11ac  & Yes & Yes & No & No & No & No & Different PHY/MAC enhancements improve TCP performance & 

  1) Consideration of both data link layer and transport layer,
  2) Trade-offs between TCP and SCTP
 & 
 1) Evaluation of SCTP with all PHY/MAC enhancements is not provided,
 2) Different network scenarios are not considered
 \\
\hline

Lee \textit{et al.} \cite{Okhwan:2015} & Energy efficient channel bandwidth management  & Yes & Yes & No & No & No & No & Helps to reduce end-to-end transmission latency & 

  1) On-demand channel bandwidth selection,
  2) Power save operation
 & 
 1) Implementation details are absent,
 2) Different channel conditions are not included,
 3) Coexistence with MU-MIMO is not explained
 \\
\hline

\end{tabular} 
\label{table:14}
\end{table*}

\begin{table*}
\caption{Comparative study of IEEE 802.11ac related works (continue...)}
\centering
\begin{tabular}{|p{1.1cm}|p{1.7cm}|l|l|p{1cm}|l|l|l|p{1.7cm}|p{2.6cm}|p{2.5cm}|}
\hline
\textbf{Item} & \textbf{Objective} & \multicolumn{6}{c|}{\textbf{Trade-off}} & \textbf{Impact on upper layer protocols} & \textbf{Strength} & \textbf{Weakness} \\
\cline{3-8} 
 & & \textbf{MU-MIMO} & \textbf{CB} & \textbf{Dynamic bandwidth management} & \textbf{SGI} & \textbf{FA} & \textbf{BA} & & & \\
\hline

Aleksandra \textit{et al.} \cite{Panajotovic:2015} & Design of MU-MIMO schemes suitable for OFDM system & Yes & No & No & No & No & No & MU-MIMO with different MCS levels helps to reduce packet transmission latency & 

  1) Finite MCS set,
  2) Uniform channel decomposition
 & 
 1) Dynamic channel management is not available,
 2) Design details of MU-MIMO are not provided
 \\
\hline

Zhiqun \textit{et al.} \cite{Zhiqun:2015} & TXOP sharing scheme of IEEE 802.11ac EDCA & Yes & No & No & No & No & No & Improves TCP/UDP performance & 

  1) Markov chains for non-saturated conditions,
  2) Downlink MU-MIMO 
 & 
 1) Considers ideal channel condition,
 2) All enhanced PHY/MAC features are not included
 \\
\hline

Juan \textit{et al.} \cite{Juan:2015} & Channel access scheme for scheduling multiuser parallel transmission & No & Yes & No & No & No & No & Channel bonding can increase both uplink and downlink application performance & 

  1) Channel bonding,
  2) Consideration of both uplink and downlink
 & 
 1) Antenna mechanism is not mentioned,
 2) Different channel conditions are not focused
 \\
\hline

Yazid \textit{et al.} \cite{Yazid:2014} & Performance analysis of TXOP sharing scheme in IEEE 802.11ac & Yes & No & No & No & No & No & Increases TCP/UDP throughput for different traffic categories & 

  1) Markov chain based analytical model,
  2) Downlink MU-MIMO, 
  3) Different access categories,
  4) Channel access fairness issue
 & 
 1) Other enhanced PHY features are not included,
 2) Design details of spatial streams are absent
 \\
\hline

Aajami \textit{et al.} \cite{Aajami:2015} & TXOP sharing based throughput-aware mechanism & Yes & No & No & No & No & No & MU-MIMO can reduce end-to-end delay & 

  1) Scheduling and joint link adaptation of downlink MU-MIMO,
  2) Ant colony optimization mechanism
 & 
 1) Explanation of included PHY features is not provided,
 2) TXOP without dynamic channel management
 \\
\hline

Narayan \textit{et al.} \cite{Narayan:2015} & Investigation of the performance differences between IEEE 802.11ac and IEEE 802.11n & Yes & Yes & No & No & Yes & No & Both standards achieve gain in TCP/UDP throughput & 

  1) Consideration of IPv4 and IPv6,
  2) Ant colony optimization mechanism
 & 
 1) Analysis of several enhanced PHY/MAC features is absent,
 2) Performance trade-offs among the features are not provided
 \\
\hline

Chung \textit{et al.} \cite{Chung:2015} & Mathematical model to analyse saturated MAC throughput & Yes & Yes & No & No & No & No & Reduces transmission delay and positively affects upper layer performance & 

  Integration of TXOP and MAC protocol data unit 
 & 
  Different enhanced PHY/MAC features are not highlighted 
 \\
\hline

Oh \textit{et al.} \cite{Oh:2014} & Performance analysis of interference alignment (IA) and MU-MIMO & Yes & No & No & No & No & No & Helps to enhance network capacity & 

  1) MU-MIMO with TDMA mechanism,
  2) Consideration of interference,
  3) Channel state information
 & 
 1) Does not consider design and implementation issues,
 2) All PHY enhancements are not included
 \\
\hline

Khavasi \textit{et al.} \cite{Khavasi:2014} & Max-min fair type link adaptation for downlink MU-MIMO & Yes & No & No & No & No & No & MU-MIMO increases rate selection capability and improves upper layer performance & 

  1) Optimal utility of max-min fair type link adaptation,
  2) Polynomial time optimal solution
 & 
 1) Interference is not explained,
 2) Only MU-MIMO for link adaptation,
 3) Upper layer performance is not analyzed
 \\
\hline

Kim \textit{et al.} \cite{Hyunsub:2014} & Low-complexity based adaptive channel state information (CSI) & Yes & Yes & No & No & No & No & Improves network architecture & 

  1) Considers interference and feedback overhead,
  2) Optimized CSI feedback
 & 

 1) Different PHY/MAC enhancements are not mentioned,
 2) Noise level is absent
 \\
\hline

\end{tabular} 
\label{table:15}
\end{table*}

\begin{table*}
\caption{Comparative study of IEEE 802.11ac related works (continue...)}
\centering
\begin{tabular}{|p{1.1cm}|p{1.7cm}|l|l|p{1cm}|l|l|l|p{1.7cm}|p{2.6cm}|p{2.5cm}|}
\hline
\textbf{Item} & \textbf{Objective} & \multicolumn{6}{c|}{\textbf{Trade-off}} & \textbf{Impact on upper layer protocols} & \textbf{Strength} & \textbf{Weakness} \\
\cline{3-8} 
 & & \textbf{MU-MIMO} & \textbf{CB} & \textbf{Dynamic bandwidth management} & \textbf{SGI} & \textbf{FA} & \textbf{BA} & & & \\
\hline

Michelle \textit{et al.} \cite{Go:2010} & Investigation of the problem of MAC in WLANs with downlink MU-MIMO & Yes & No & No & No & No & No & MU-MIMO with beamforming reduces end-to-end delay & 

 1) Proposes a MAC protocol for downlink MU MIMO,
 2) Beamforming based approach,
 3) Considers hidden node problem
 & 
 1) Implementation details are absent,
 2) All PHY/MAC enhanced features are not included
 \\
\hline

Oscar \textit{et al.} \cite{Oscar:2013} & Important enhanced features of IEEE 802.11ac & Yes & Yes & Yes & No & No & No & Shows the improvement of several network performance factors which influence TCP/UDP throughput & 

Focuses on channel bonding and MU-MIMO 
 & 
 Trade-offs among different PHY/MAC enhanced features are not discussed\\
\hline

Aleksandra \textit{et al.} \cite{Panajotovic:2013} & MU-MIMO based OFDM for IEEE 802.11ac & Yes & No & No & No & No & No & MU-MIMO can help to improve network capacity & 

  Different decompositions of channel matrix
 & 
 1) Synchronization and design of MU-MIMO antenna are not discussed,
 2) Channel bonding is not focused
 \\
\hline

Mounir \textit{et al.} \cite{Esslaoui:2012} & Multiuser fair scheduling scheme based on MU-MIMO OFDM & Yes & No & No & No & No & No & Helps to appropriately select different transmission parameters and increase user level throughput & 

  1) Simultaneous transmissions to several users,
  2) Appropriate transmission parameters for selected user
 & 
 1) Interference with noise is not considered,
 2) Fairness without channel bonding
 \\
\hline

\end{tabular} 
\label{table:16}
\end{table*}






\begin{table*}
\caption{IEEE 802.11 related works which can be introduced in IEEE 802.11n/ac}
\centering
\begin{tabular}{|p{1cm}|p{5cm}|p{3.5cm}|p{6.9cm}|}
\hline
\textbf{Item} & \textbf{Strength} & \textbf{Impact on upper layer protocols} & \textbf{Possible Future Extensions}\\
\hline

Manshaei \textit{et al.} \cite{Manshaei:2009} & 

 1) Identifies the key parameters needed to design a rate adaptation mechanism,
 2) Designs three IEEE 802.11 rate adaptation algorithms 
 & Identification of key rate adaptation parameters increases TCP/UDP throughput &
 1) Performance evaluation of different rate adaptation algorithms under different IEEE 802.11n/ac enhanced features, 
 2) Evaluates the new proposed algorithms with IEEE 802.11n/ac features
 \\
\hline

Xia \textit{et al.} \cite{Xia:2008} & Designs a rate adaptation technique utilizing statistics with Received Signal Strength Indicator to maximize the throughput & Enhances TCP/UDP performance & Performance evaluation of Smart Sender algorithm under different IEEE 802.11n/ac enhanced features\\
\hline

Dong \textit{et al.} \cite{Dong:2013} & 

 1) Evaluation of Minstrel algorithm in IEEE 802.11g, 
 2) Loopholes in Minstrel 
 & Improves end-to-end throughput & 
 1) Performance evaluation of Minstrel in IEEE 802.11n and IEEE 802.11ac,
 2) Possible required enhancements in Minstrel for better adjustment in dynamic channel condition
 \\
\hline

Tahiry \textit{et al.} \cite{Tahiry:2006} & Solves the performance anomaly problem in IEEE 802.11 using packet aggregation and dynamic time slot & Helps to reduce packet transmission delay & 

 1) Performance evaluation of the proposed approach in IEEE 802.11n/ac, 
 2) Consideration of all PHY/MAC enhancements in the proposed scheme
 \\
\hline

Osama \textit{et al.} \cite{Amr:2012} & Designs a multi-band scheme for dynamically switching between $2.4$ GHz and $5$ GHz bands & Dynamic channel switching can reduce packet loss and helps to positively affect TCP/UDP throughput &

 1) Impact of this proposed scheme over application level throughput,
 2) Performance evaluation of this scheme under different IEEE 802.11n/ac enhanced parameters
\\
\hline

Biaz \textit{et al.} \cite{Biaz:2008} & Presents a comprehensive study of the advances of rate adaptation mechanisms for IEEE 802.11 & Increases TCP/UDP performance & Designs a rate adaptation algorithm for high throughput WLAN using all the enhanced features of IEEE 802.11n and IEEE 802.11ac\\
\hline
\end{tabular} 
\label{table:17}
\end{table*}

\end{document}